%% file: DN-18-007_paper.tex
\documentclass[11pt,a4paper]{article}
\pdfoutput=1
\usepackage{jinstpub}
\usepackage{color}
\usepackage{lineno,hyperref,xspace,topcapt}
\modulolinenumbers[5]

\title{The effect of radiation damage on the light yield and
  uniformity of candidate plastic scintillator tiles for the CMS
  hadron calorimeter upgrade}

\collaboration{The CMS HCAL Collaboration}
\newcommand{\unit}[1]{\ensuremath{\text{\,#1}}\xspace}
\newcommand{\avgpe}{\ensuremath{\langle\mathrm{PE}\rangle}\xspace}

\newcommand{\GeV}{\ensuremath{\text{\,Ge\hspace{-.08em}V}}\xspace}
\newcommand{\fbinv} {\mbox{\ensuremath{\,\text{fb}^{-1}}}\xspace}

\abstract{
\input{content/abstract.tex}
}


\arxivnumber{2207.11960}

\begin{document}

\maketitle

\flushbottom

%%%%%%%%%%%%%%%

\input{content/main_text.tex}

%%%%%%%%%%%%%%%

\acknowledgments
\input{content/acknowledgements.tex}

\bibliography{DN-18-007_paper}

\clearpage
\section*{The CMS HCAL Collaboration}
\addcontentsline{toc}{section}{The CMS HCAL Collaboration}
\input{DN-18-007_authorlist_optC}

\end{document}

%% file: content/abstract.tex
A study has been performed to understand the effects of radiation
damage on various plastic scintillator tiles considered for a possible
upgrade of the hadron calorimeter of the CMS
detector. Measurements were made with unirradiated tiles and with
tiles that had been irradiated in the CMS collision hall to a dose of
44\unit{kGy}. Results are presented for the tiles of different shapes
in terms of the energy spectrum, efficiency as a function of the
position at which each tile was hit, as well as light yield. All the
tiles showed a light reduction of up to about 50\%. The tiles with the
shape currently used in the CMS detector did not see increased
non-uniformity of light collection, while a significant disuniformity
was observed for the tiles considered as alternatives.

%% file: content/main_text.tex
\section{Introduction}
\label{sec:introduction}
Modern large-scale particle physics experiments at high
center-of-mass energies, such as ATLAS~\cite{ATLAS:2008xda} and
CMS~\cite{CMS:2008xjf} at the Large Hadron Collider (LHC) facility at
CERN, use plastic scintillators for particle detection due to their
low cost and large light yield. However, scintillators are subject to
radiation damage, reducing their light
yield~\cite{Zorn:1992qn,osti_22979632} over the time scale of the
experiment. For example, the scintillating tiles in the endcap
hadronic calorimeter (HE) of the CMS detector had a reduction of light
output up to about 25\% after the 2017 LHC operation, which
corresponded to an integrated luminosity of
50\fbinv~\cite{CMS:2020mce} of proton-proton collisions, and a dose of
approximately 1.5\unit{kGy}. It is anticipated that the scintillators
in the hadron calorimeters for the proposed experiments at the future
accelerators, such as FCC-hh, could receive doses up to 8\unit{kGy} or
even 1\unit{MGy}~\cite{Aleksa:2019pvl}. Studies of the radiation
effects on the transmission of light in plastic scintillators have
been presented, for example, in Ref.~\cite{Liao:2015zsa}. This study
presents a continuation of the attempt to find a solution to the
decreased light yield of the plastic scintillators in the CMS hadronic
endcap calorimeter, which covers the rapidity region of
${1.4<|\eta|<3}$, where ${\eta = -\ln\left(\tan\frac{\theta}{2}\right)}$
and $\theta$ is the polar angle, measured with respect to the beam line.

Our previous paper~\cite{CMSHCAL:2017gwh} compared the performance of
the tiles made from scintillator materials manufactured by the Eljen
corporation\footnote{Eljen Technology, 1300 W. Broadway, Sweetwater,
  TX 79556, United States} (EJ-200, EJ-260, with different dopant
concentrations) to those used in the legacy calorimeter (SCSN-81 by
Kuraray\footnote{Kuraray, Ote Center Building,1-1-3, Otemachi,
  Chiyoda-ku, Tokyo, 100-8115, Japan }) of the CMS detector; the
production of Kuraray has since been discontinued. It is believed that
the effects of radiation damage can be partially mitigated by either
overdoping the material or by shifting the output to a longer
wavelength region. The previous study confirmed that both of these
were viable options for the replacement of the SCSN-81 tiles. We now
present an assessment of how these materials perform when damaged by
radiation. The irradiation was performed at the Castor Radiation
Facility at CERN with the scintillator samples installed in the
proximity of the LHC beam line. In addition to the two previous
methods of mitigating the effects of radiation damage, a third
possibility, using tiles of different shapes, is also checked. The
light yield of the scintillator detectors is measured in terms of
their response to minimum ionizing particles.

%St. Gobain Corp.\footnote{Saint Gobain Corp, Les Miroirs, 18, Avenue
%d'Alsace, 92400 Courbevoie, France}

\section{Experimental setup}
\label{sec:experimental_setup}

\begin{figure}[!h]
  \begin{center}
    \includegraphics[width=0.9\textwidth]{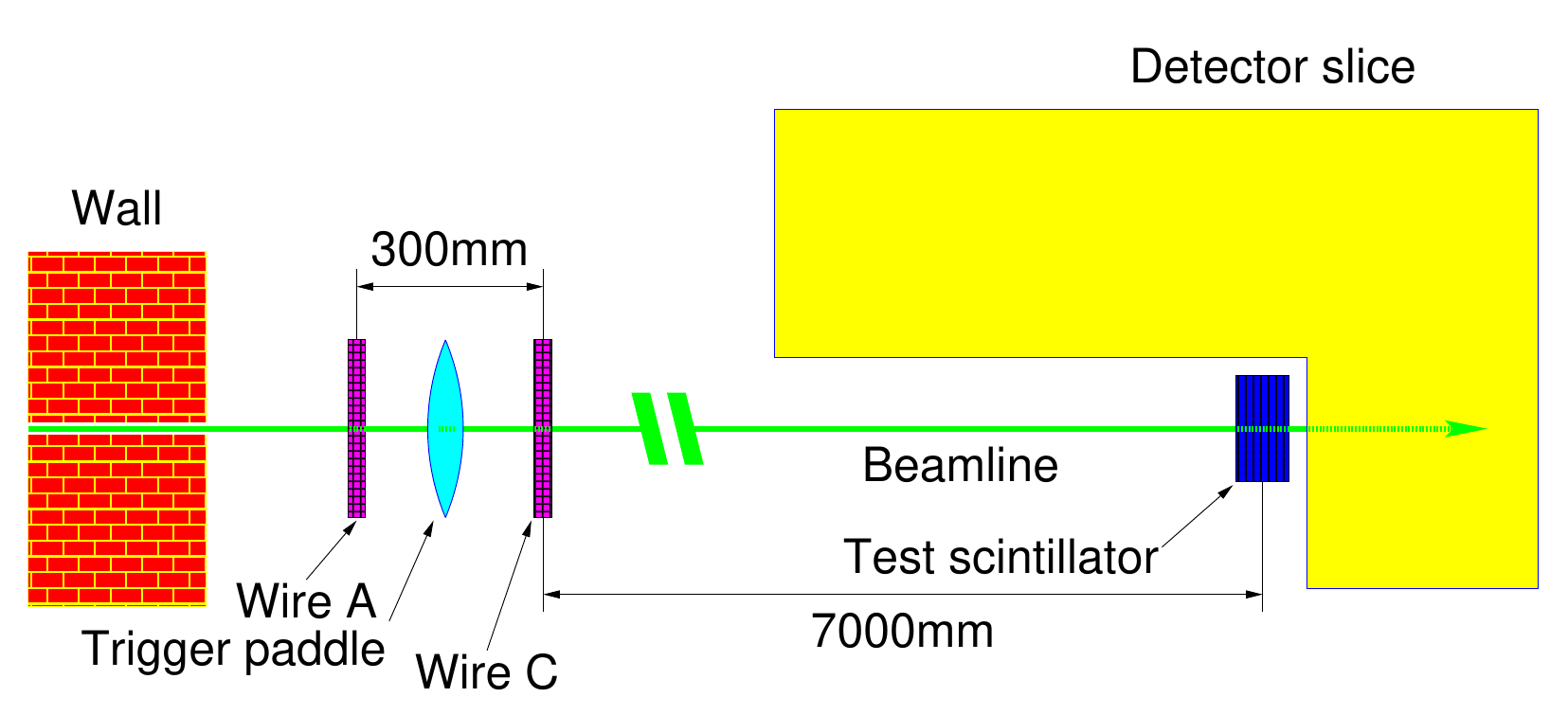}
    \caption{Diagram of the experimental area, not to scale.}
    \label{fig:exp_area}
  \end{center}
\end{figure}

The experimental setup is almost identical to the one in our previous
study~\cite{CMSHCAL:2017gwh}. The data were collected in the North
Area of the H2 Beam Line at the CERN Prevessin site using beam dumps
from the Super Proton Synchrotron (SPS). A muon beam is generated from
the decay of a 150\GeV charged pion beam. The beam path passes through
several instruments for triggering and precise tracking on its way to
the scintillator samples under test. In order of increasing distance
from the beam extraction point, these are wire chamber ``A'', four
large scintillating plastic trigger counters, and wire chamber
``C''. The wire chambers were the drift chambers described in
Ref.~\cite{CMSHCAL:2017gwh}, with size ($100\times100\unit{mm}^2$) and
resolutions of $\approx 0.5\unit{mm}$ in the transverse plane.
Figure~\ref{fig:exp_area} shows the relative positions of the
instruments, tiles, and beam. Only two of the wire chambers were
active for this study, as opposed to the three used in
Ref.~\cite{CMSHCAL:2017gwh}. The yellow region indicates an
instrumented $40^{\circ}$ wedge of HCAL, which is not used in the
analysis presented in this paper.

% The area_diagram.pdf file is produced with the following command:
%
% > fig2dev -L pdf -F figs/area_diagram.fig| \
%   sed s/"%PDF-1.7"/"%PDF-1.5"/ > figs/area_diagram.pdf
%
% Note the forced change of PDF minor version: tdr warnings if >5

\begin{figure}[!h]
  \begin{center}
    \includegraphics[width=0.485\textwidth]{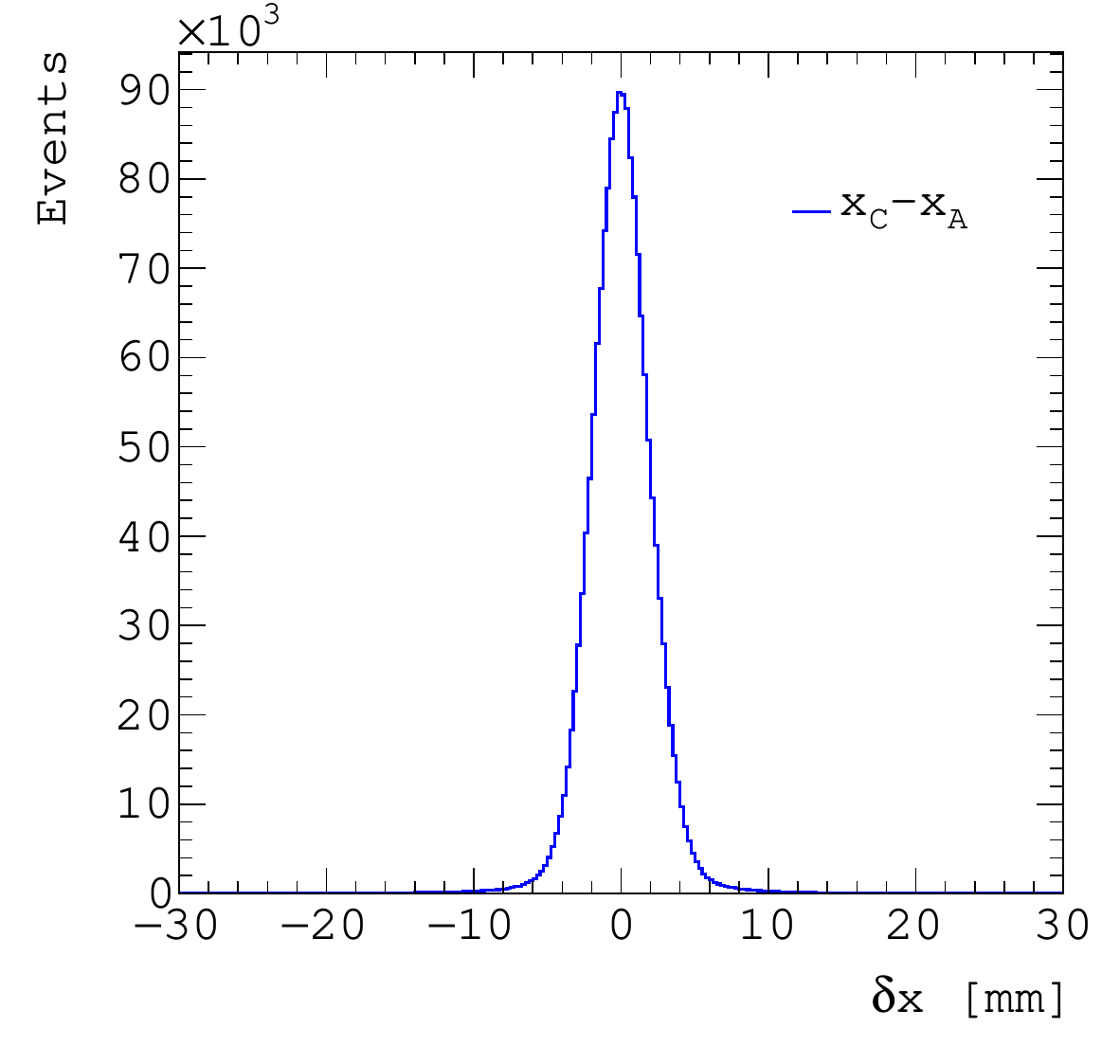}%
    \includegraphics[width=0.485\textwidth]{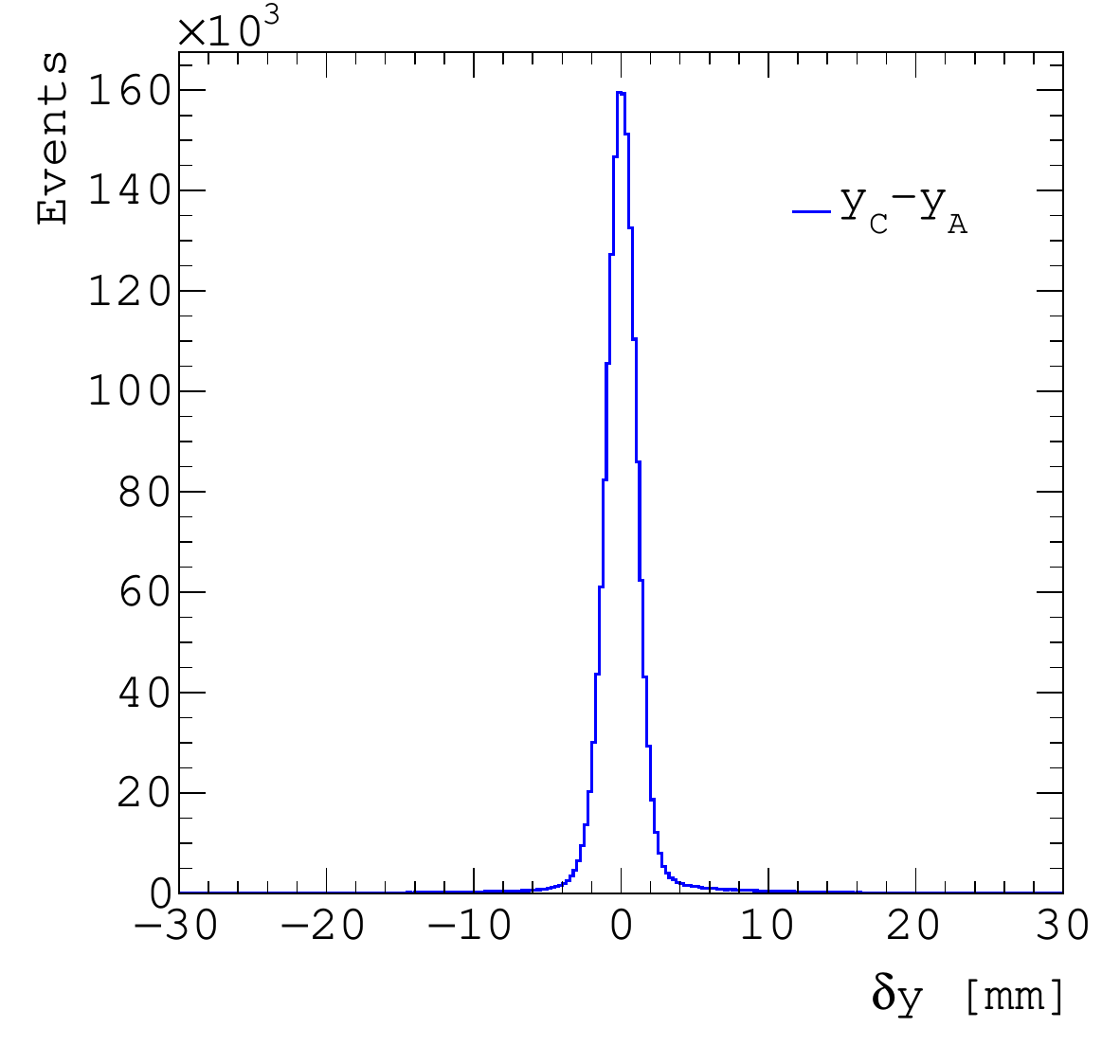}
    \caption{Difference in position of hits along the $x$ (left) and $y$
    (right) directions for the two wire chambers. The
    distributions contain an offset correction calculated assuming
    that the muons travel along straight trajectories.}
    \label{fig:deltaxy}
  \end{center}
\end{figure}

The two wire chambers were aligned before use in the analysis. The
hits in the wire chambers determine the points where the muons from
the incident beam intersect the tiles. We assume that the muons in the
experimental area travel along a straight line, and therefore the
difference between the $x$ and $y$ measurements made in the two wire
chambers is a Gaussian with mean zero. The distributions of the $x$
and $y$ measurements after the alignment are shown in
Fig.~\ref{fig:deltaxy}.

\begin{figure}[!h]
  \begin{center}
    \includegraphics[width=0.9\textwidth]{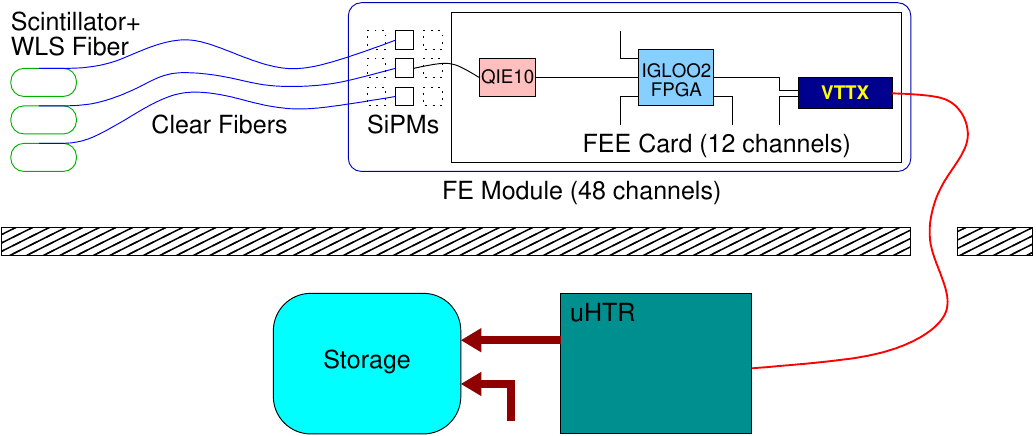}
    \caption{Overview of the data path at the H2 test beam facility at CERN.}
    \label{fig:daq}
  \end{center}
\end{figure}

The data acquisition system has front-end and back-end electronics
designed for the Phase-I upgrade of the CMS hadron calorimeter (HCAL)
with silicon photomultipliers (SiPMs) as the photodetectors. The
system is described in detail in Ref.~\cite{CMS:2012tda}. The
experimental setup is based on the Phase-I configuration of
SiPM-equipped HCAL detectors, as reported in
Ref.~\protect\cite[Fig.~1.7]{CMS:2012tda}. Each scintillator tile is
connected to an individual SiPM (although in the CMS detector each
SiPM receives scintillation light from multiple tiles). The light
produced in the scintillator is transmitted via wavelength-shifting
(WLS) fibers to silicon photomultiplier (SiPM) photodetectors. The
SiPMs are custom-made by Hamamatsu Photonics\footnote{Hamamatsu
  Photonics, 325-6, Sunayama-cho, Naka-ku, Hamamatsu City, Shizuoka,
  430-8587, Japan} to fulfill the requirements for their usage in the
HCAL subsystem of the CMS experiment. A detailed description of their
characteristics is presented in Ref.~\cite{Heering:2016aqi}. The pulse
of current generated by the SiPM is integrated using
Charge-Integrator-and-Encoder (QIE10)
chips~\cite{Zimmerman:2003gv,Baumbaugh:2014eya,Roy:2015bia,Hare:2016swd},
with the digitized signals collected by a Microsemi IGLOO2 FPGA and
transmitted to the back-end electronics via an optical link through a
versatile-link transmitter (VTTx)~\cite{Vasey:2012xjw}. The VTTx
connects each front-end (FE) module to an HCAL $\mu$TCA Trigger and
Readout ($\mu$HTR) module. The encoded signals from the QIE10 chips
and the wire chambers are reconstructed into events and saved to disk.
Figure~\ref{fig:daq} indicates the path of the signal from the
scintillator tiles to the storage via the digitization (QIE10), the
alignment and formatting (Microsemi IGLOO2 FPGA), the data
transmission (VTTx), the back-end electronics (uHTR).

\begin{figure}[!h]
  \begin{center}
    \raisebox{-0.5\height}{\includegraphics[height=7cm]{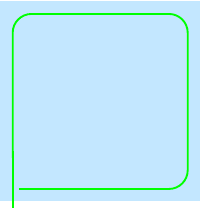}}%
    \hspace*{1cm}
    \raisebox{-0.5\height}{\includegraphics[height=7.1cm]{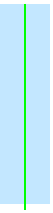}}%
    \hspace*{1cm}
    \raisebox{-0.5\height}{\includegraphics[width=0.33\textwidth]{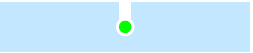}}
    \caption{Schematic drawing of the $\sigma$ (left) and finger
    (center) tiles and a sectional view of the groove within which the
    WLS fiber is installed.}
    \label{fig:tile_formats}
  \end{center}
\end{figure}

The scintillator tiles had two types of grooves where WLS fibers are
inserted to collect a fraction of the light produced in the tile. We
refer to them as $\sigma$ and {\it finger} tiles, respectively. A
schematic drawing of the two formats is sketched in
Fig.~\ref{fig:tile_formats}. The same figure provides a sectional view
of the tile showing the groove within which the WLS fiber is
installed. In the first case the WLS fiber roughly follows a
$\sigma$-shaped path along the border of the $\sigma$ tile, while for
the second it goes straight through the middle of the finger tile. The
$\sigma$ tiles have areas of $10\times 10\unit{cm}^2$, while the
finger tiles are $10\times 2\unit{cm}^2$. The thickness of the tiles
were the same, about $0.4\unit{cm}$.

% The pdf files are produced with a similar command to the one used
% for the area_diagram.pdf file

A summary of the characteristics of the scintillator tiles used in
this experiment is given in Table~\ref{tab:scintillators}. The Eljen
scintillator tiles were exclusively available in $\sigma$-tile format,
while SCSN-81 tiles were available in both $\sigma$- and finger-tile
formats. The WLS fiber is Y11 for blue scintillators and O2 for green
scintillators, both produced by Kuraray.

\begin{table}[!h]
  \begin{center}
  \topcaption{Basic characteristics of the scintillator tiles used in
    this study. The tile color refers to the wavelength of maximum
    emission, i.e., $425\unit{nm}$ for the blue scintillators and
    $490\unit{nm}$ for the green. The irradiation was performed at the
    Castor Radiation Facility at CERN with the scintillator samples
    installed in the proximity of the LHC beam line. The base material
    of the Eljen (EJ) tiles is polyvinyl toluene (PVT), while
    polystyrene (PS) is used for the SCSN tiles, produced by
    Kuraray. The EJ tile marked by ``$\dagger$'' contains a
    proprietary primary dopant, in a concentration about twice than
    available in commercial samples.}
  \label{tab:scintillators}
    \begin{tabular}{ l | c | c | c | c}
      Tile      & Base & Color & Format [$\unit{cm}^3$]& Integrated dose [$\unit{kGy}$]\\
      \hline
      EJ 200    & PVT & Blue  &  $10\times 10\times 0.4$ & not irradiated\\
      EJ 260    & PVT & Green &  $10\times 10\times 0.4$ & not irradiated\\
      EJ 260 2P$^\dagger$
                & PVT & Green &  $10\times 10\times 0.4$ & not irradiated\\
      SCSN 81F1 & PS  & Blue  &  $2\times 10\times 0.37$ & $44\pm 7$\\
      SCSN 81F2 & PS  & Blue  &  $2\times 10\times 0.37$ & $44\pm 7$\\
      SCSN 81F3 & PS  & Blue  &  $2\times 10\times 0.37$ & $44\pm 7$\\
      SCSN 81F4 & PS  & Blue  &  $2\times 10\times 0.37$ & $44\pm 7$\\
      SCSN 81S  & PS  & Blue  & $10\times 10\times 0.37$ & $55\pm 9$\\
    \end{tabular}
  \end{center}
\end{table}

The SCSN-81 tiles were subjected to irradiation before being included
in the test-beam measurement. They were placed in the CMS collision
hall on the structure that housed the CMS forward calorimeter
CASTOR~\cite{Gottlicher:2010zz}, 14.3\unit{m} away from the CMS
interaction point, and at a distance between $13.1$ and
$14.4\unit{cm}$ from the LHC beam line. The total integrated dose was
measured using FWT-60-00 Radiachromic dosimeters (thin films) by Far
West Technology, which were attached to the top of the scintillator
samples. The uncertainty in the radiation dose is estimated to be
$15\%$.

\section{Data analysis}
\label{sec:data_sample}

Prior to any selection, the data consisted of about 7 million
triggered events. Events are required to have a single muon hitting
both wire chambers. This ensures that the energy corresponds to a
single minimum-ionizing particle (MIP) signal and allows the
measurement of the detection efficiency as a function of the hit
position on a tile. This requirement reduces the event sample to
1,720,742 events for the subsequent analysis. The finger tiles were
not used until a later stage of the data taking operation; hence the
corresponding sample contained only about 2.8 million events, which
reduced to just 622,635 events after applying the requirement of
single-muon traversal.

\begin{figure}[!h]
  \begin{center}
    \includegraphics[width=0.485\textwidth]{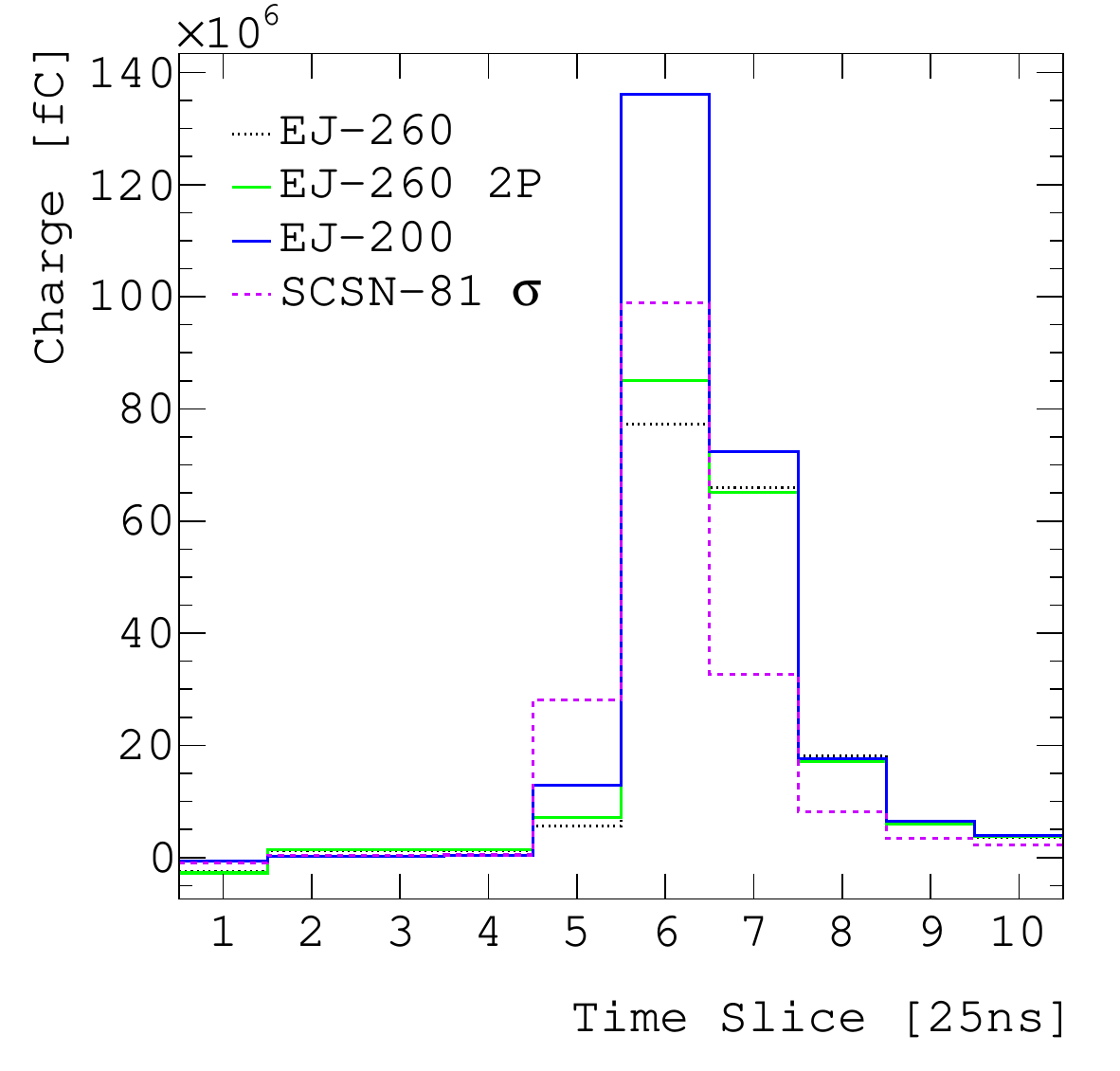}%
    \includegraphics[width=0.485\textwidth]{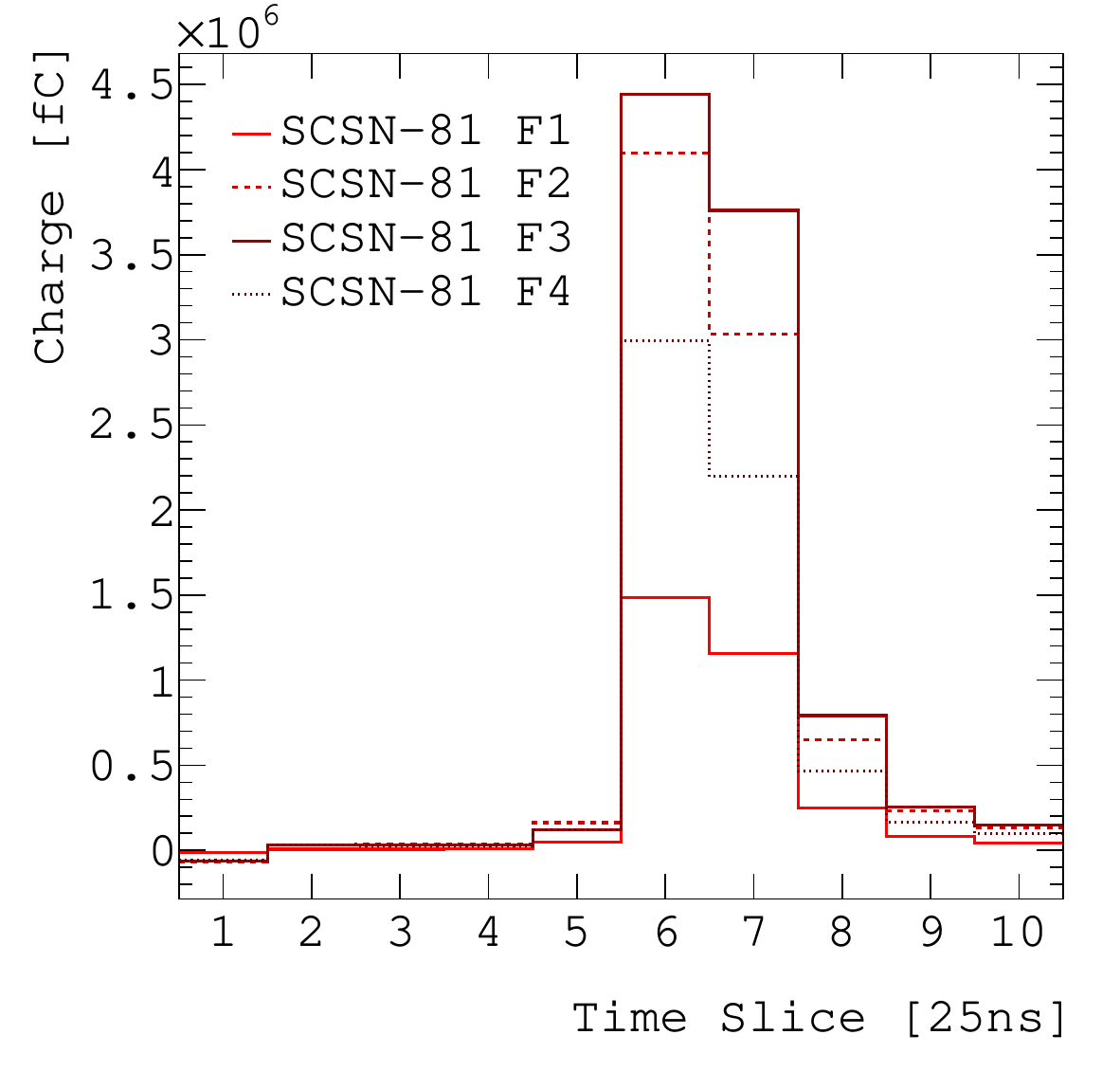}
    \caption{Average integrated SiPM charge in 25\unit{ns} time slices
    for $\sigma$ tiles (left) and finger tiles (right) in a muon
    beam.}
    \label{fig:ts}
  \end{center}
\end{figure}

The data acquisition system measures a charge, corresponding to the
time-integrated current from the SiPM photosensors in 25\unit{ns} time
slices. The integrated charge is proportional to the number of
photoelectrons, which in turn is proportional to the energy deposited
by a MIP that crosses the scintillator. Figure~\ref{fig:ts} shows the
measured pedestal-subtracted average charge in each of ten time slices
for the finger tiles and for the $\sigma$ tiles. Time slices 5
through 9 are used in the subsequent analysis to maximize the signal
to noise ratio. From these plots it can be seen that all tiles were
timed-in similarly and the same time slices are therefore used for
each tile.

The event-by-event per-time-slice pedestal is estimated by averaging
the charge integrated in the first three time slices. Unless
otherwise noted, the event-by-event pedestal is estimated by
multiplying the per-time-slice pedestal value estimated above by the
number of time slices that are added to obtain the total integrated
charge in the event, i.e., time slices 5 through 9.

%An example of the pedestal spectrum measured using SiPMs is
%presented in Fig.~\ref{fig:energy_nofid}. The scintillator used to
%obtain the spectrum corresponds to the first SCSN-81 finger tile. No
%cutoff is applied on the number or position of incident muons, and the
%distribution is dominated by pedestal noise, centered around the value
%of zero integrated over charge that extends up to about 25\unit{fC}. The
%dark current and electronic noise are among the main contributors to
%the pedestal peak. Light leakage is minimized by careful wrapping of
%the scintillator tiles. The distribution in logarithmic scale presents
%a set of equally spaced Gaussian peaks, each corresponding to the
%collection of an increasing number of photoelectrons (\pe).
%This distribution demonstrates that 25\unit{fC} provides
%an adequate threshold to identify events with no measured photoelectrons.

%\begin{figure}[!h]
%  \begin{center}
%    \includegraphics[width=0.485\textwidth]{figs/energy_PS_bins_pref_SCSN_81F1.pdf}%
%    \includegraphics[width=0.485\textwidth]{figs/energy_PS_bins_pref_log_SCSN_81F1.pdf}
%    \caption{Pedestal distribution for the SCSN-81 F1 tile, before any
%    requirement is set on the number or position of incident muons.
%    The dashed line at 25\unit{fC} marks the threshold used to distinguish
%    energy depositions associated with a muon hit from noise. Each
%    Gaussian peak corresponds to the collection of an increasing
%    number of photoelectrons. The same distribution is shown in both
%    the linear and logarithmic scales.}
%    \label{fig:energy_nofid}
%  \end{center}
%\end{figure}

\subsection{Hit efficiency}
\label{sec:hit_eff_measurement}
\begin{figure}[!h]
  \begin{center}
    \includegraphics[width=0.485\textwidth]{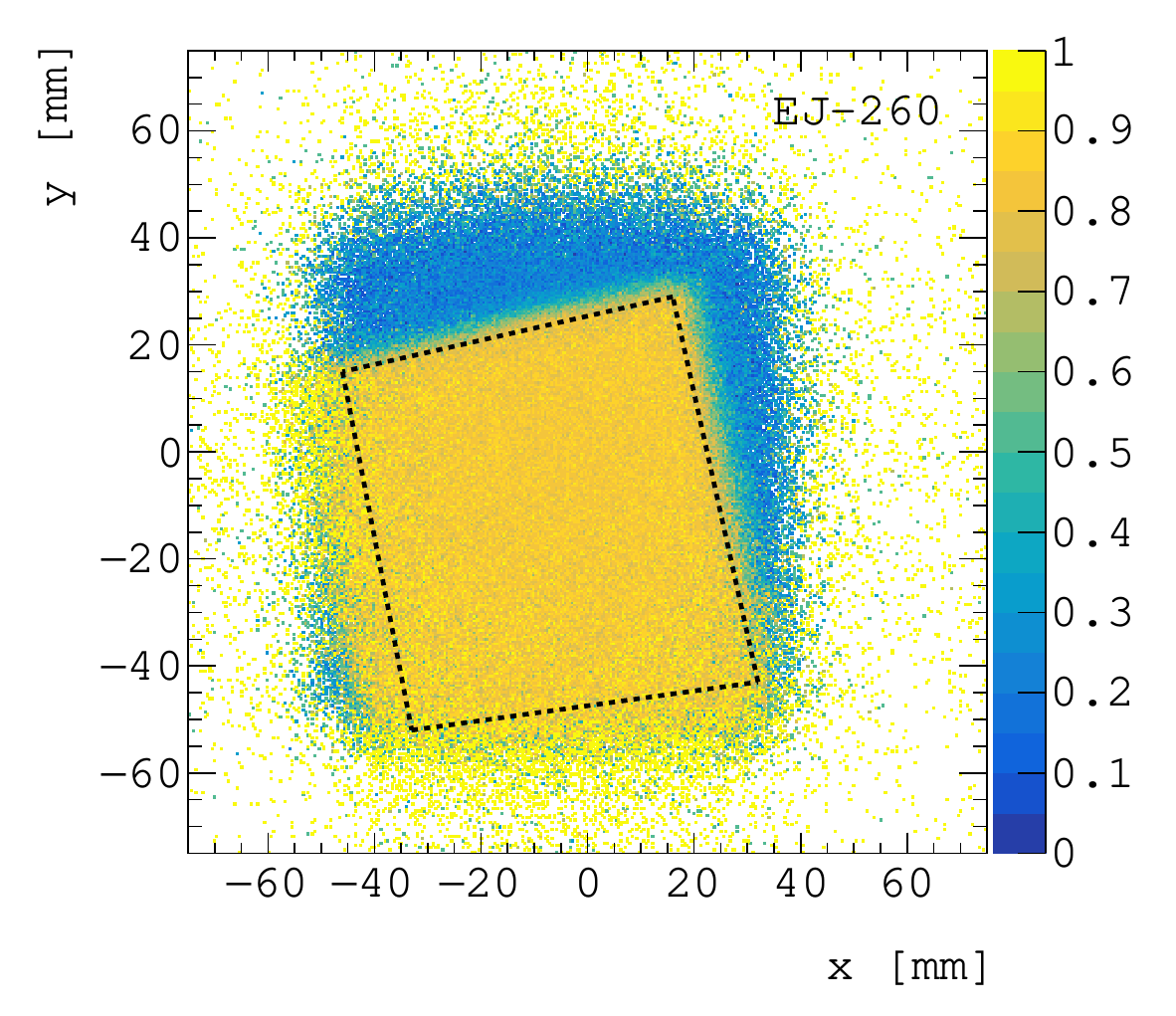}%
    \includegraphics[width=0.485\textwidth]{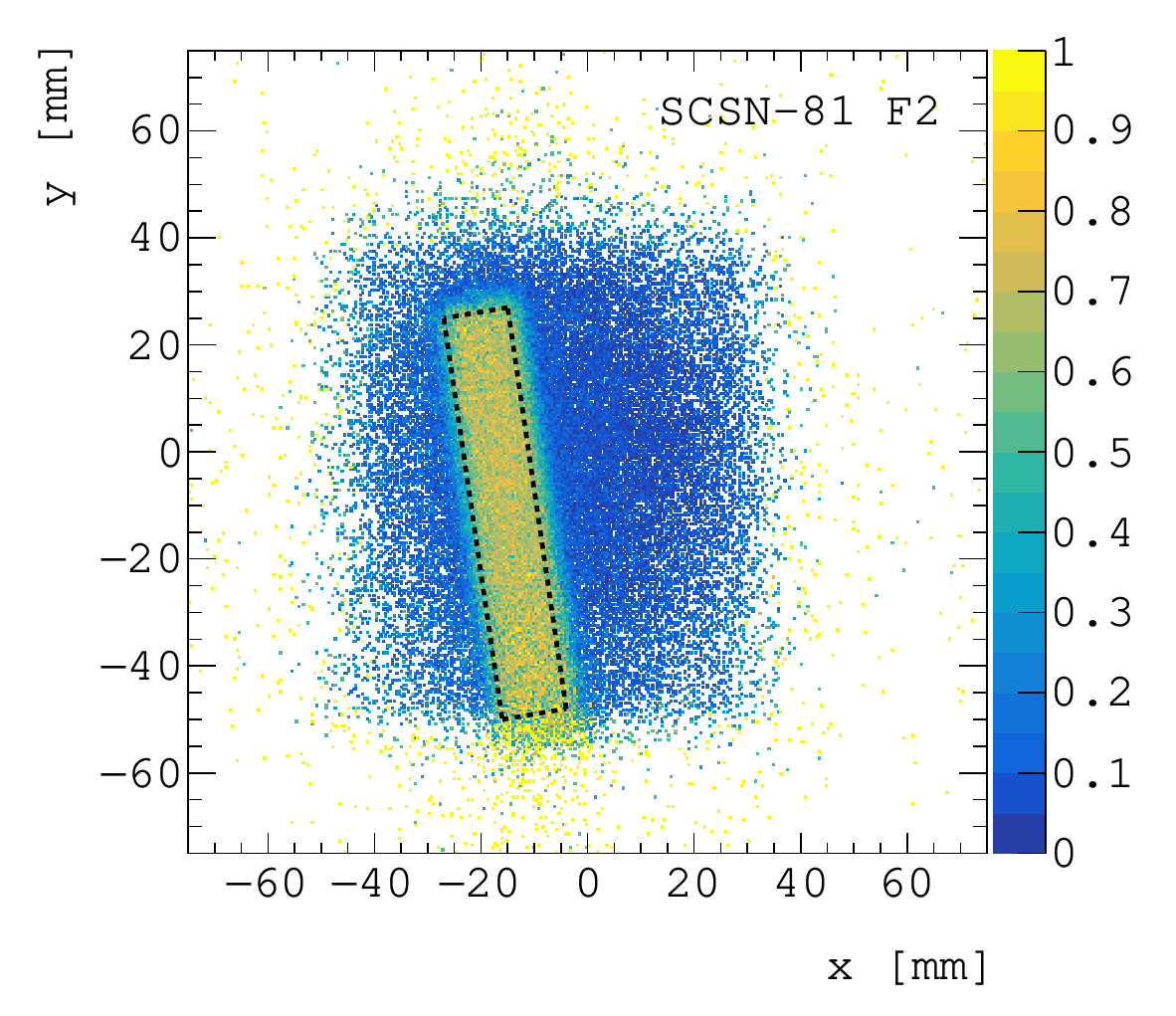}
    \caption{Maps of 2D efficiency for EJ-260 (left) and the second
    SCSN-81 finger tile (right). The dashed lines indicate the
    fiducial region that corresponds approximately to the overlap
    between the tile area and the beam, positioned along a path in
    which the hit efficiency, determined from the ratio of hits with
    an integrated pulse $>25\unit{fC}$, divided by all the hits in the
    same bin, is $>50\%$. Similar maps have been produced for the
    other scintillator samples.}
  \label{fig:fiducial}
  \end{center}
\end{figure}

Each hit is represented by a two-dimensional (2D) point denoted by $x$
and $y$ coordinates in a plane perpendicular to the beam. The hit
efficiency is defined to be the ratio of the number of events recorded
with an integrated pulse greater than 25\unit{fC}, which is the value
that separates the pedestal peak from the signal peak corresponding to
one photoelectron, to the total number of recorded events that satisfy
the single-muon traversal requirement. Two-dimensional efficiency maps
are generated for each scintillator. The efficiency maps are also used
to define a fiducial region, identifying the position of each tile
relative to the beam spot. The lines marking the fiducial region are
drawn through the region where the efficiency appeared to be at least
$\approx 50\%$, and correspond to the estimated position of the
physical boundary of the tiles. Figure~\ref{fig:fiducial} shows the
efficiency maps for the EJ-260 and the second SCSN-81 finger tile,
where the fiducial regions are identified by the dashed lines.

%Noise is studied by requiring that there be no muons within the
%fiducial region. The integrated charge distribution
%obtained is shown in Fig.~\ref{fig:afid}. The result
%also checks that %the pedestal subtraction is properly performed in
%each channel: each entry contains the integrated charge in time
%slices 5 to 9, minus the event-by-event pedestal; the distribution
%thus shows that the pedestal estimate obtained using the first three
%time slices is consistent with the energy measured by the SiPM, in
%the time-slice range of interest, when no muon is crossing
%the %corresponding tile.

%\begin{figure}[!h]
%  \begin{center}
%    \includegraphics[width=0.485\textwidth]{figs/energyPS_all_afid.pdf}
%    \caption{Integrated charge spectra produced by requiring no muon
%    crossing the %tile under study. The agreement within a factor of
%    two in the distributions for all tiles indicates that there was
%    no major difference in the sensors and signal paths of each
%    tile.}
%    \label{fig:afid}
%  \end{center}
%\end{figure}

\begin{figure}[!h]
  \begin{center}
    \includegraphics[width=0.485\textwidth]{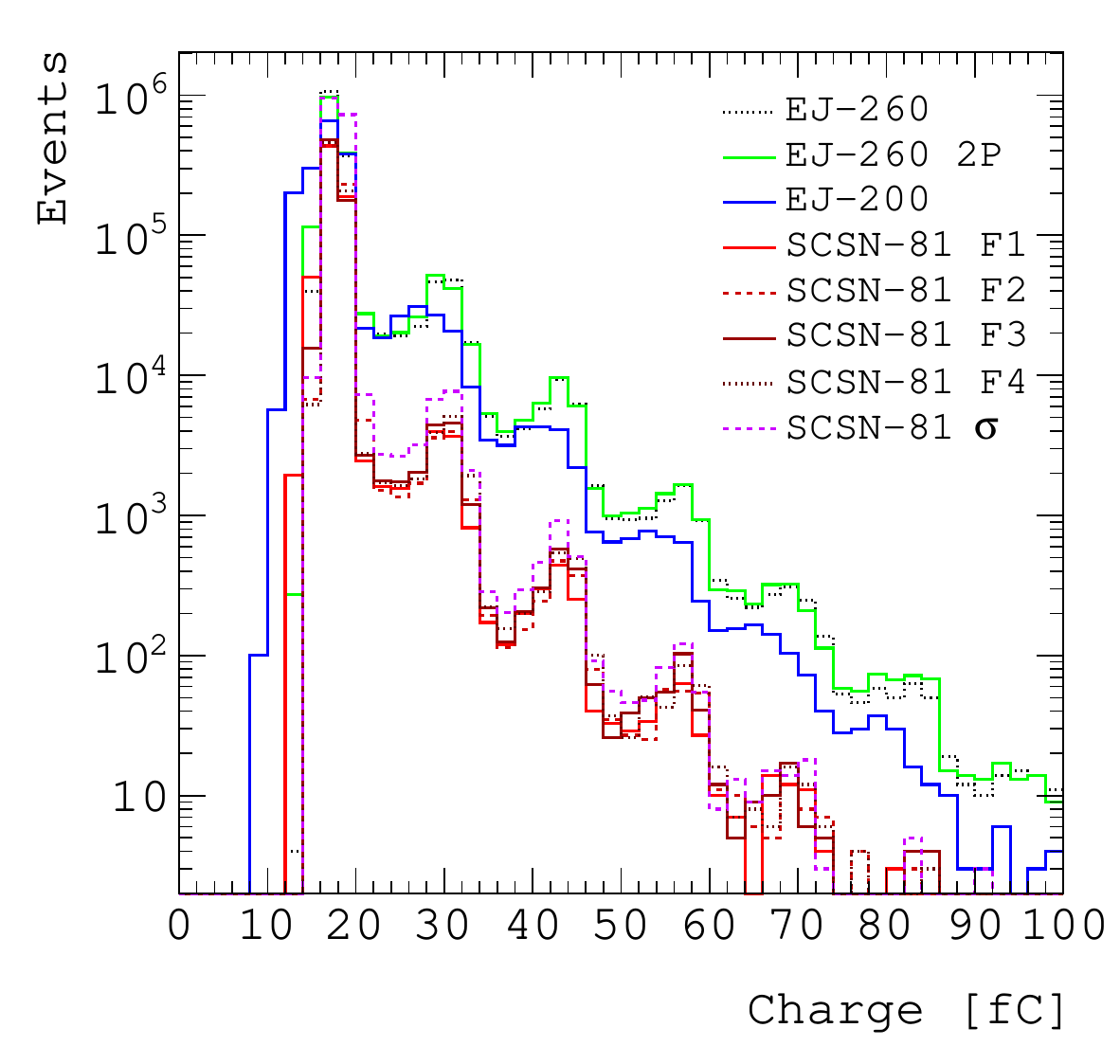}
    \caption{Event-by-event measurement of pedestal integrated charge,
    normalized to a single time slice. The distributions follow two
    distinct trends, which correspond to different SiPM arrays to
    which the scintillator tiles are connected.}
    \label{fig:pedestal}
  \end{center}
\end{figure}

The distributions in the pedestal levels, normalized to a single time
slice, are shown in Fig.~\ref{fig:pedestal}. They appear to follow two
distinct trends, which are likely due to a different amount of light
leaking through the connections, as the two sets of SiPMs are attached
to different SiPM arrays by two separate fiber bundles. The equally
spaced peaks correspond to an increasing number of photoelectrons.
It is appropriate to note that the spacing between peaks is consistent
among all SiPMs, which indicates that they have similar particle-detection
efficiencies.

% Here we will show all of the rotated efficiency plots, as well as
% the stretched out energy spectra, and the X and Y efficiency plots
\subsection{Energy spectra and efficiency maps}
\label{sec:energy_eff_maps}

\begin{figure}[!h]
  \begin{center}
    \includegraphics[width=0.485\textwidth]{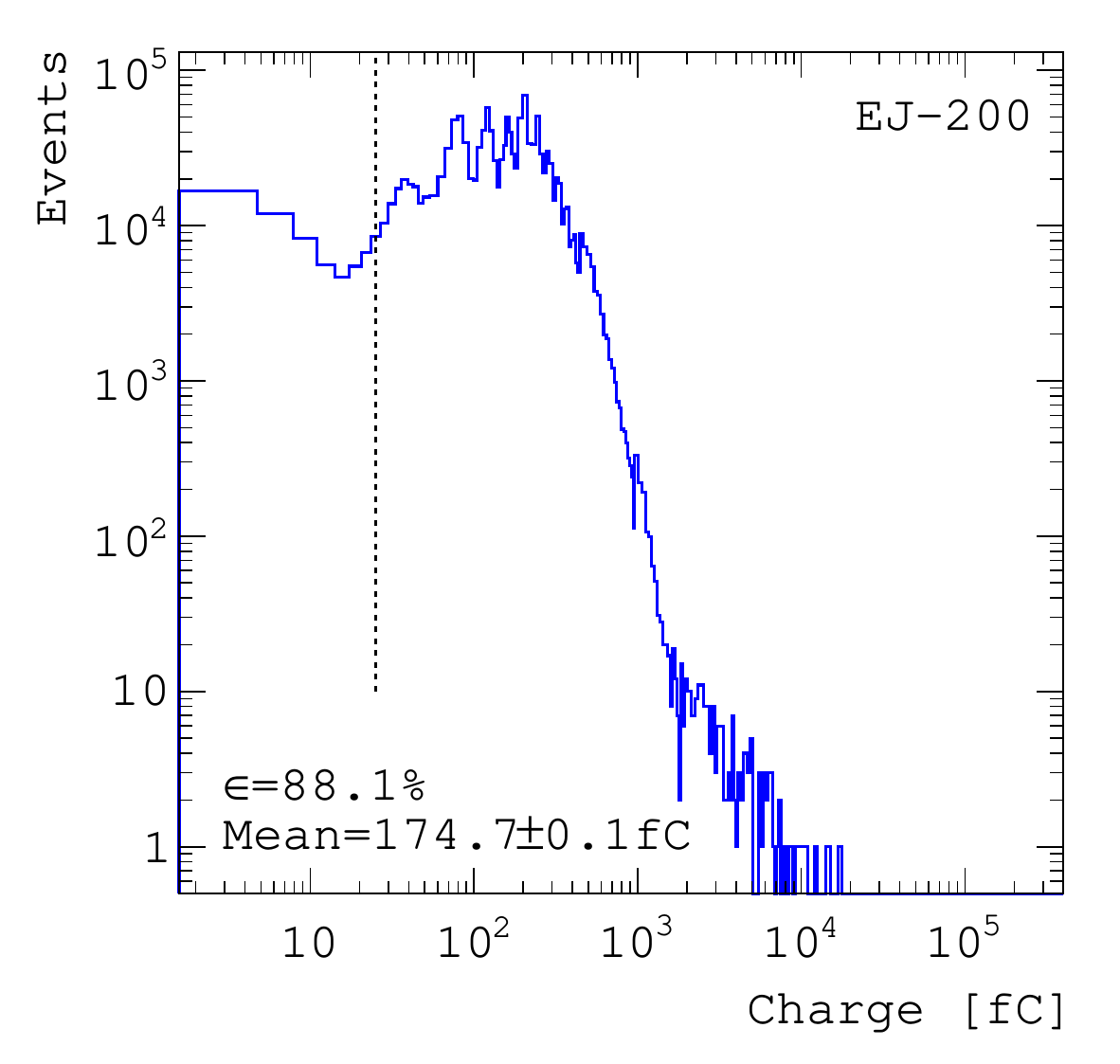}%
    \includegraphics[width=0.485\textwidth]{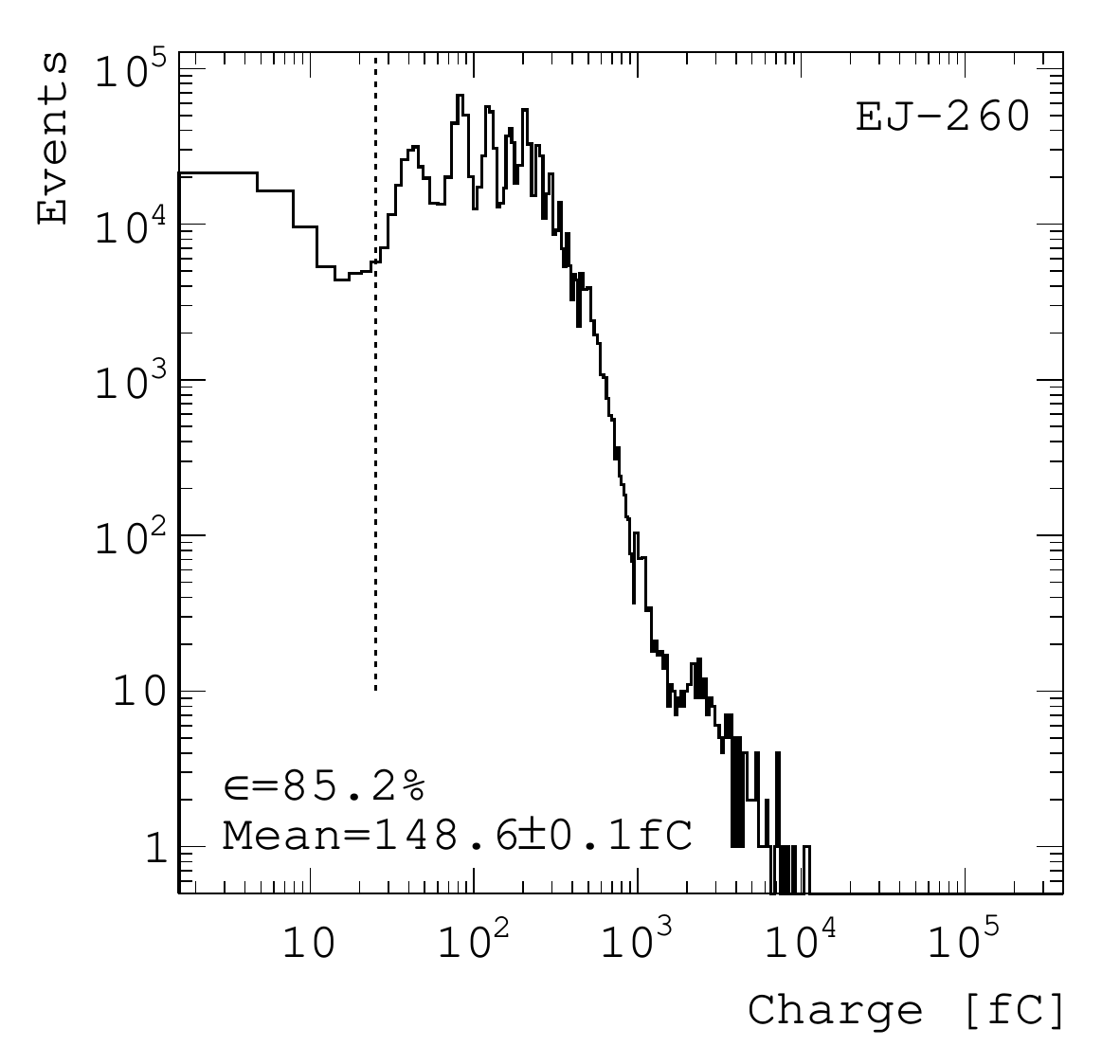}
    \includegraphics[width=0.485\textwidth]{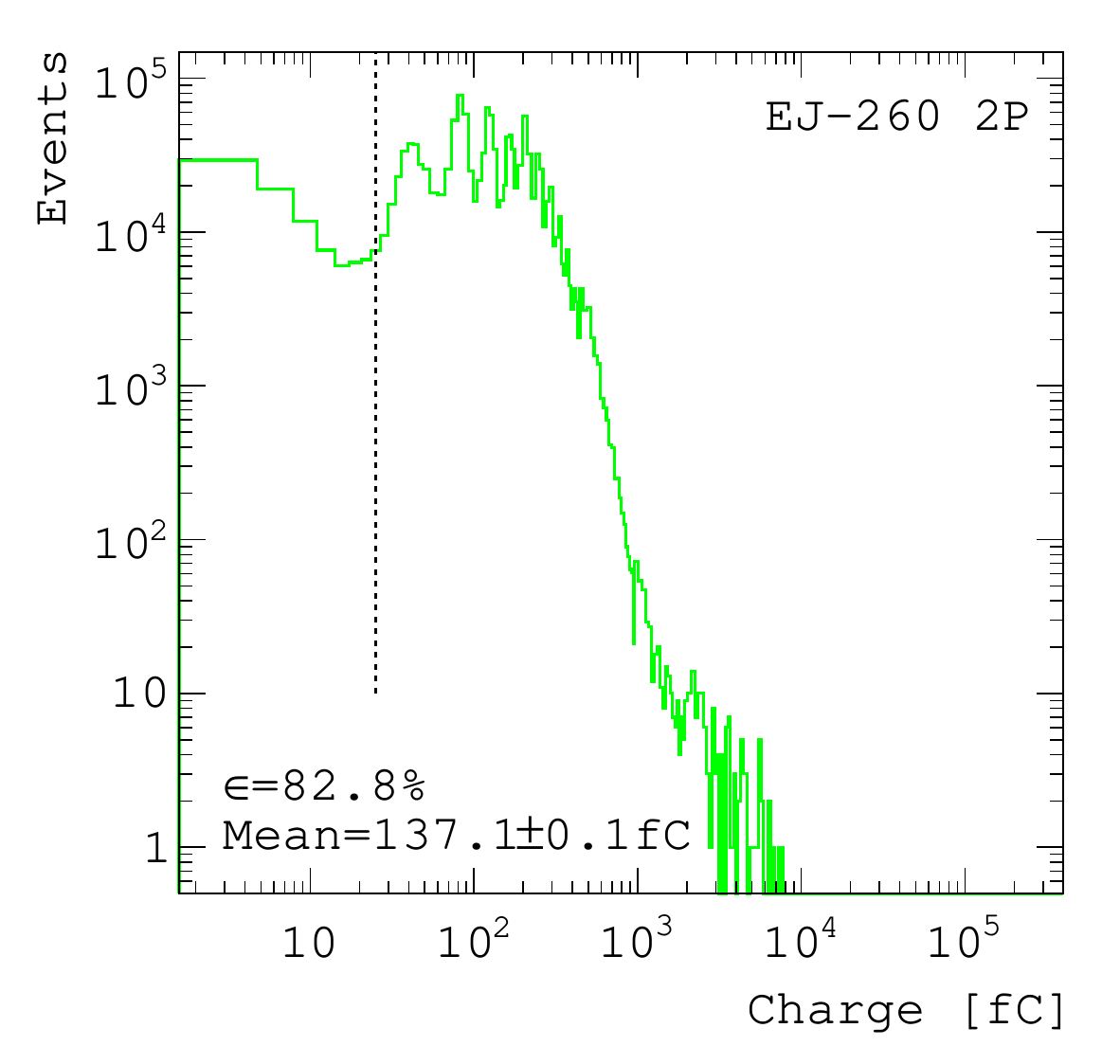}%
    \includegraphics[width=0.485\textwidth]{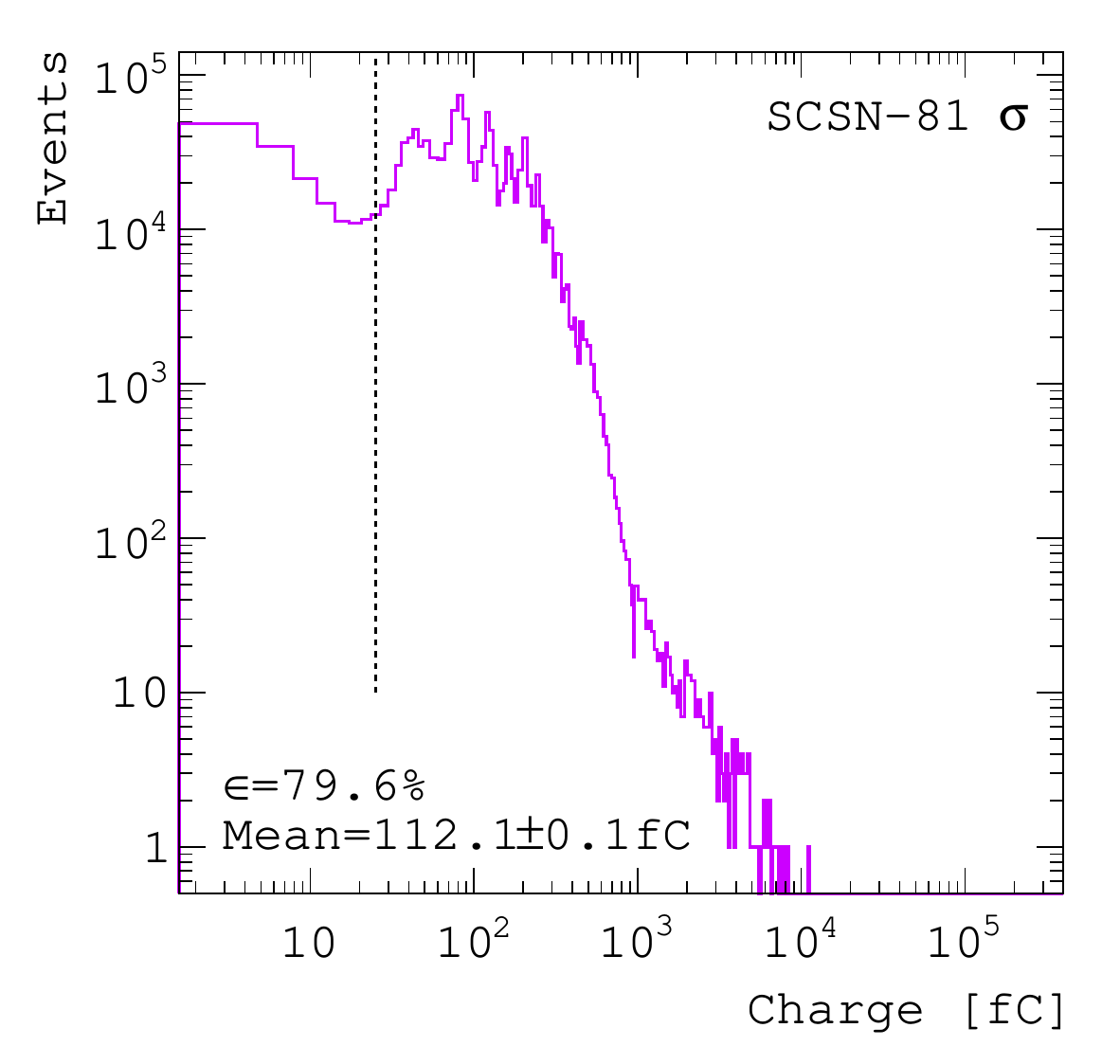}
    \caption{Integrated charge spectra for the $\sigma$ tiles
    studied. The dashed lines indicate the threshold, set at
    25\unit{fC}, above which a hit is considered to correspond to a
    MIP (muon).}
    \label{fig:energy_sigma}
  \end{center}
\end{figure}

\begin{figure}[!h]
  \begin{center}
    \includegraphics[width=0.485\textwidth]{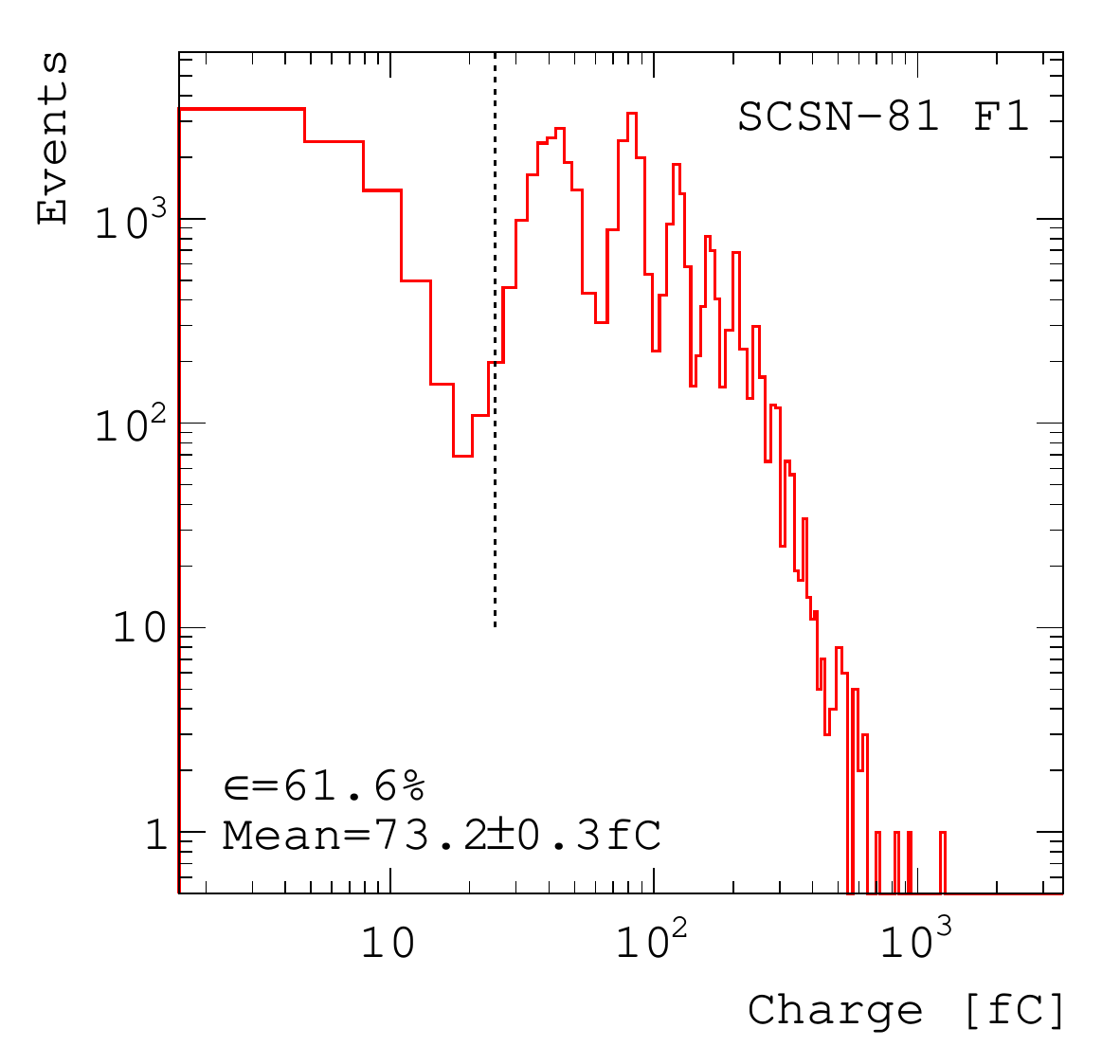}%
    \includegraphics[width=0.485\textwidth]{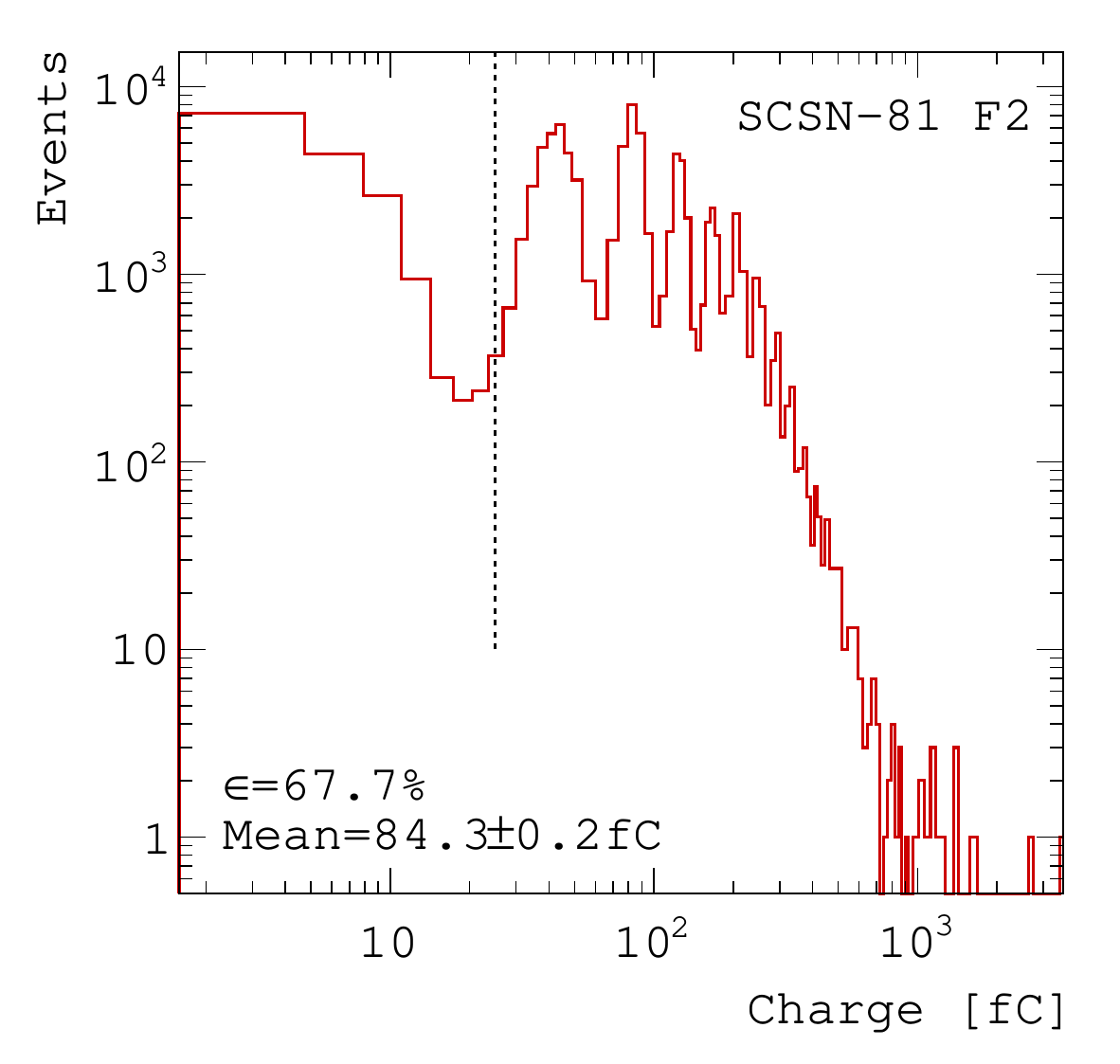}
    \includegraphics[width=0.485\textwidth]{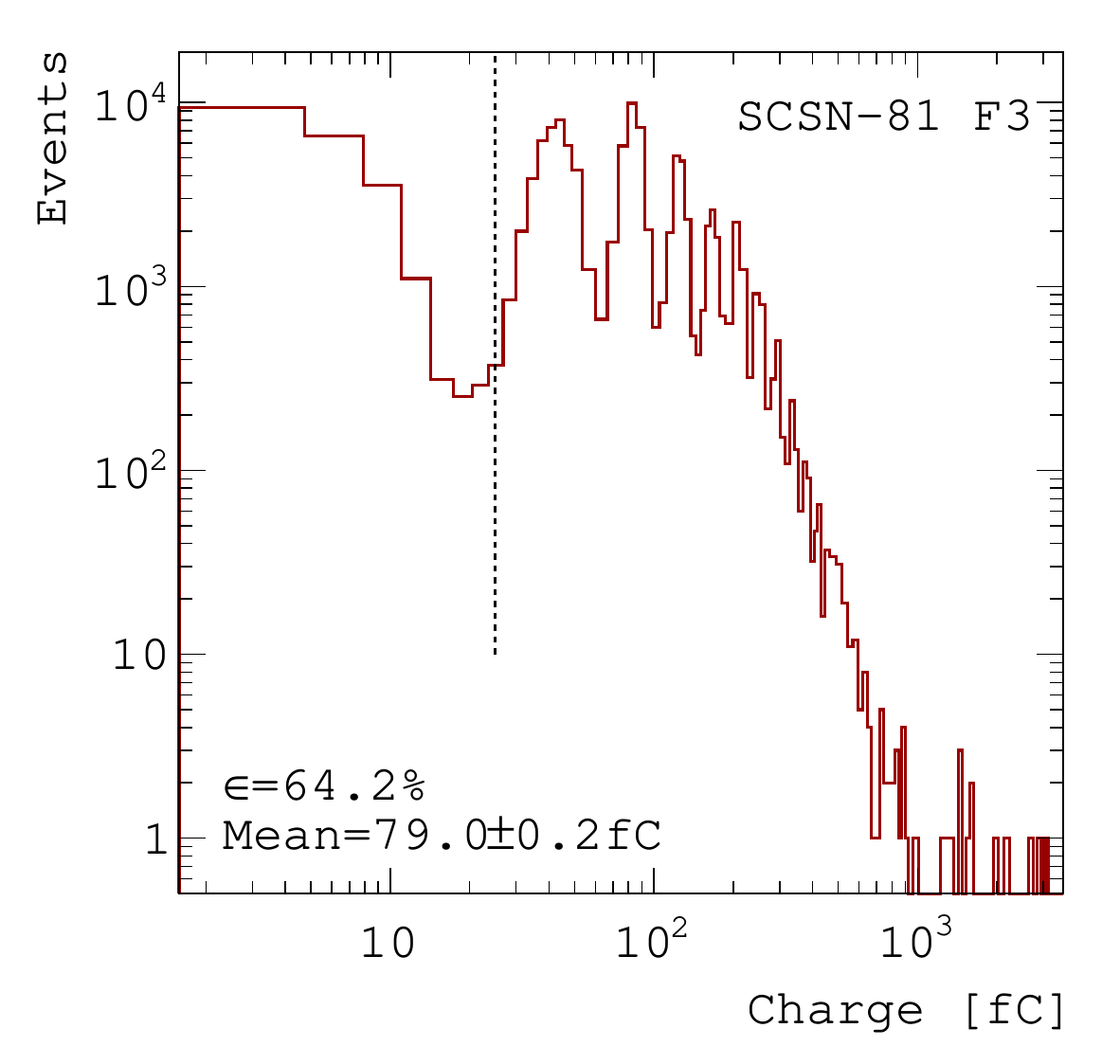}%
    \includegraphics[width=0.485\textwidth]{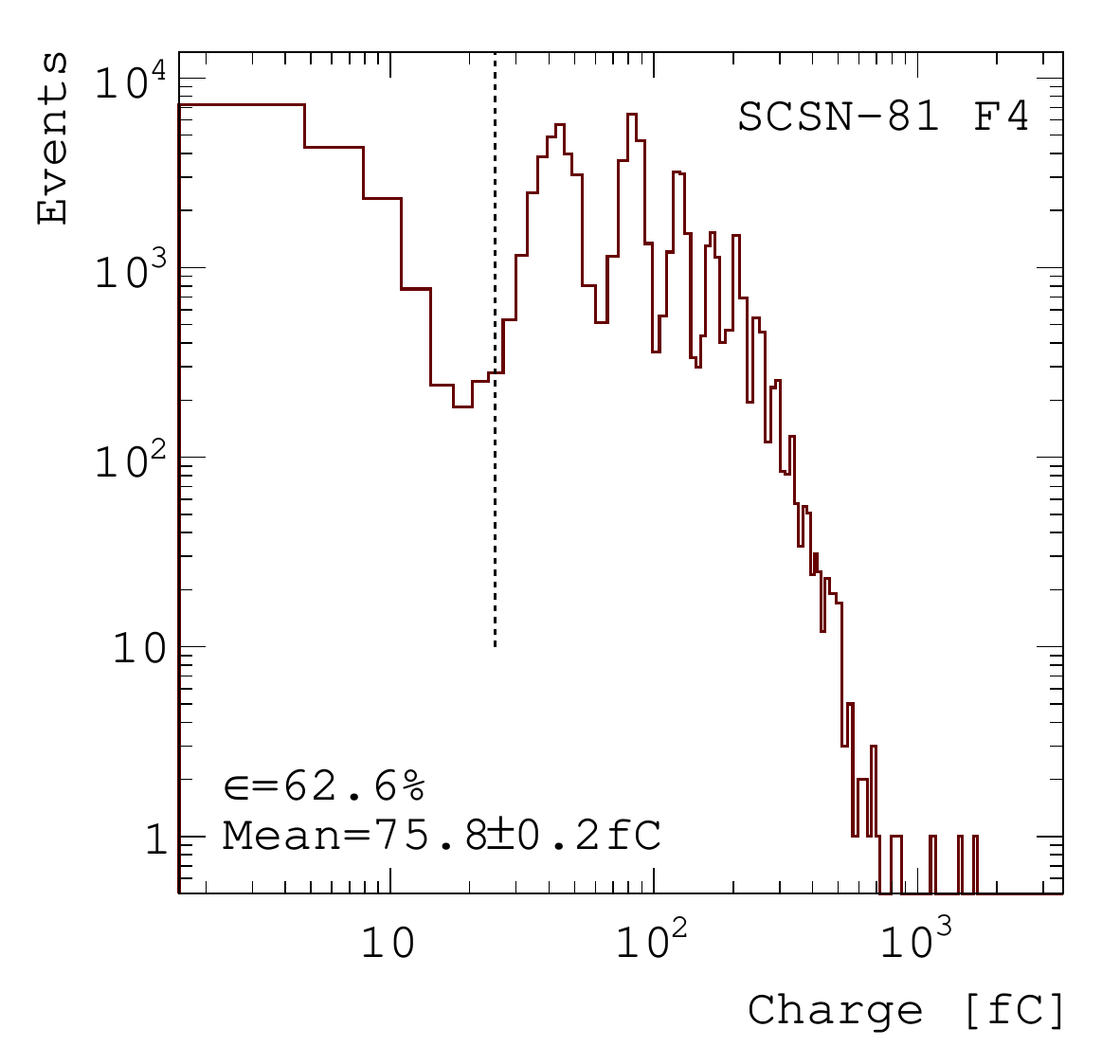}
    \caption{Integrated charge spectra for the finger tiles
    studied. The dashed lines indicate the threshold, set at
    25\unit{fC}, above which a hit is considered to correspond to a
    MIP (muon).}
    \label{fig:energy_finger}
  \end{center}
\end{figure}

The fiducial regions defined in the previous sections are used to
obtain a more accurate picture of the amount of energy collected by
the tile from each muon that interacted with the plastic
scintillator. Unless stated to the contrary, it is assumed that each
event is selected by requiring the presence of a single muon within
the fiducial region of each tile. Figures~\ref{fig:energy_sigma}
and~\ref{fig:energy_finger} show the integrated charge spectra for
fiducial muons in all of the tiles used in our study. The efficiency
$\epsilon$, introduced in Section~\ref{sec:hit_eff_measurement}, is
defined to be the ratio of the number of hits with an integrated pulse
above the noise threshold, set at 25\unit{fC}, divided by the total
number of hits.

\begin{figure}[!h]
  \begin{center}
    \includegraphics[width=0.485\textwidth]{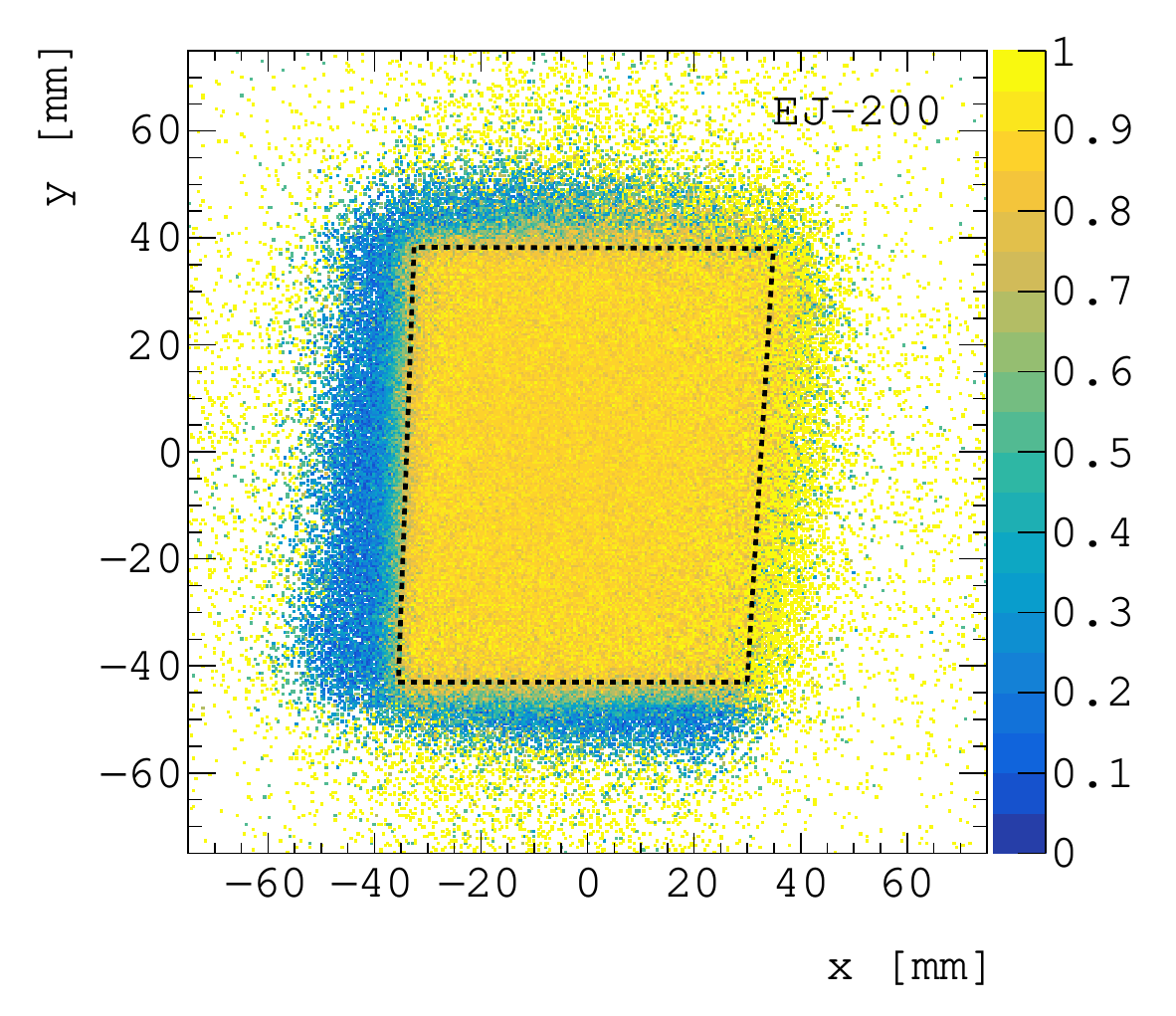}%
    \includegraphics[width=0.485\textwidth]{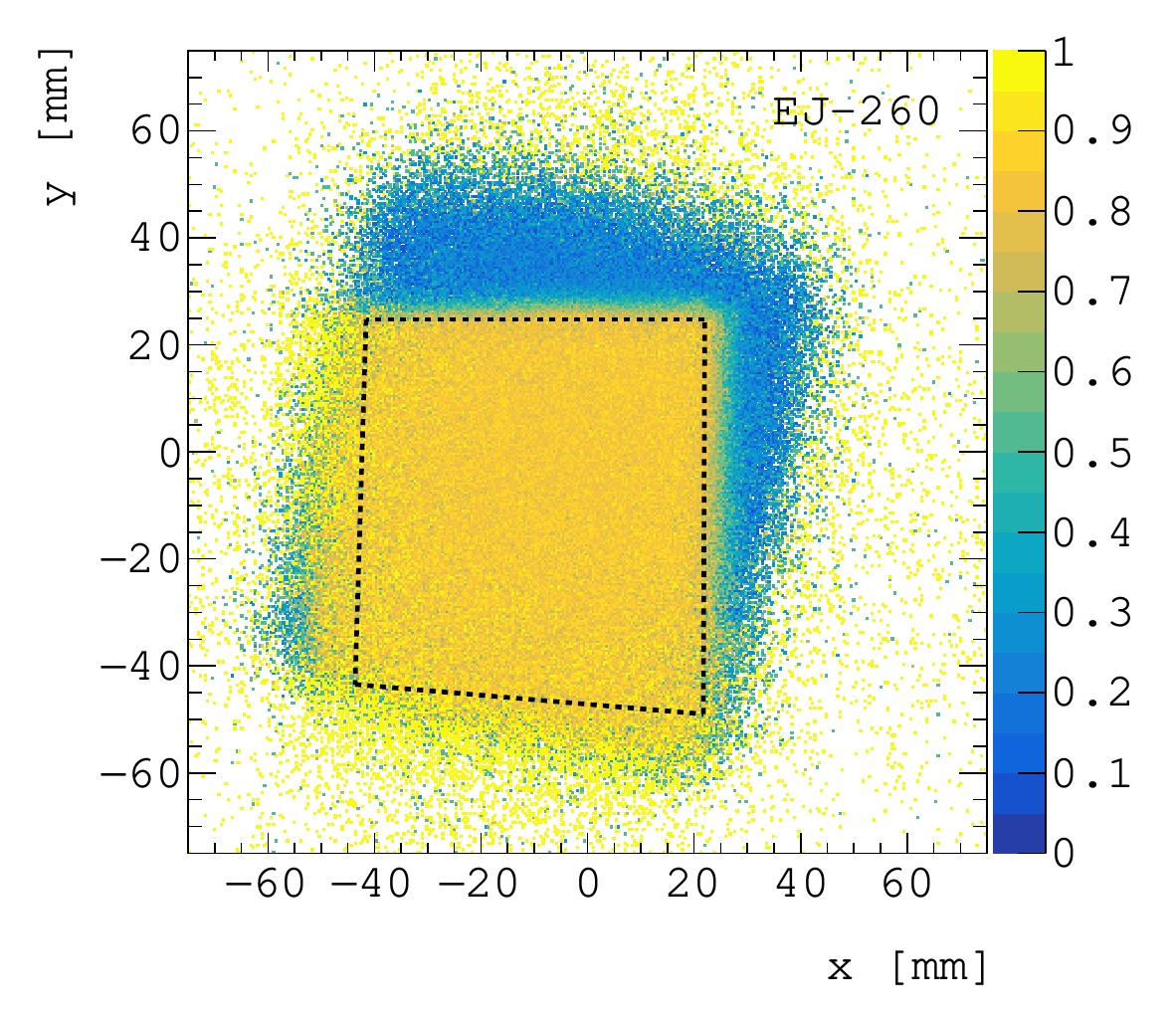}
    \includegraphics[width=0.485\textwidth]{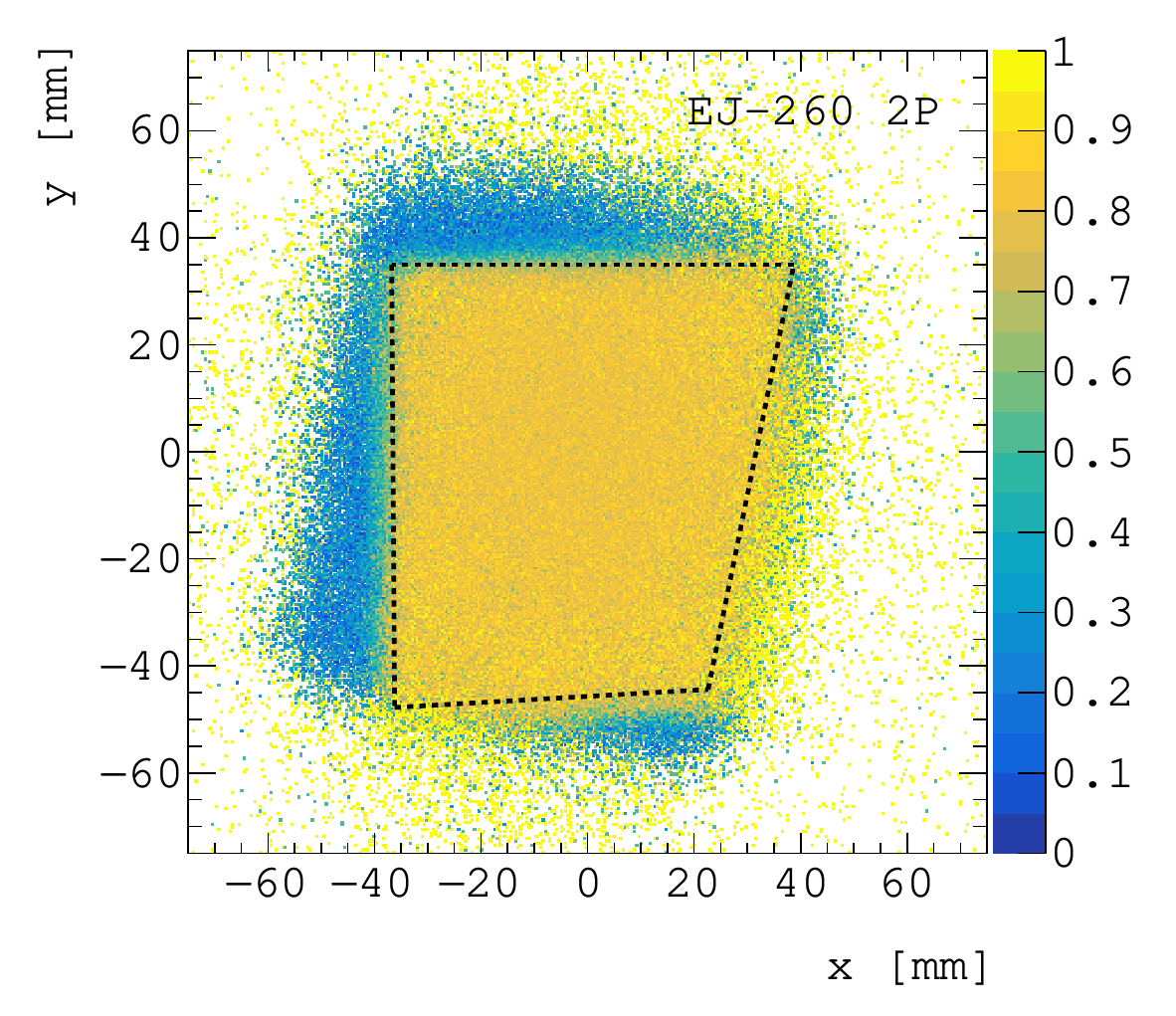}%
    \includegraphics[width=0.485\textwidth]{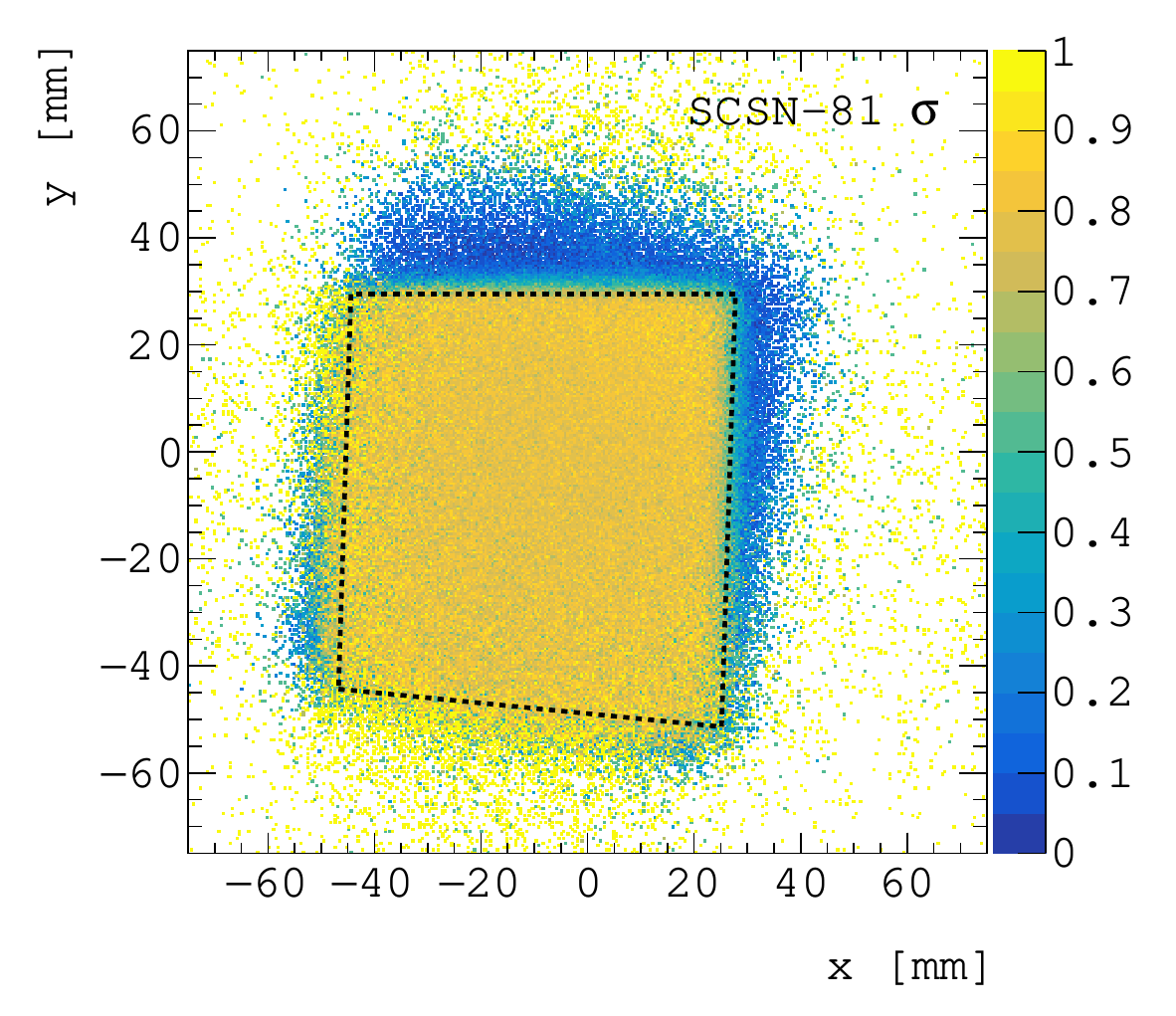}
    \caption{Efficiency maps for the $\sigma$ tiles studied. The
      dashed lines indicate the fiducial region that corresponds
      approximately to the overlap between the tile area and the beam,
      positioned along a path in which the hit efficiency, determined
      from the ratio of hits with an integrated pulse $>25\unit{fC}$,
      divided by all the hits in the same bin, is $>50\%$.}
    \label{fig:maps_sigma}
  \end{center}
\end{figure}

\begin{figure}[!h]
  \begin{center}
    \includegraphics[width=0.485\textwidth]{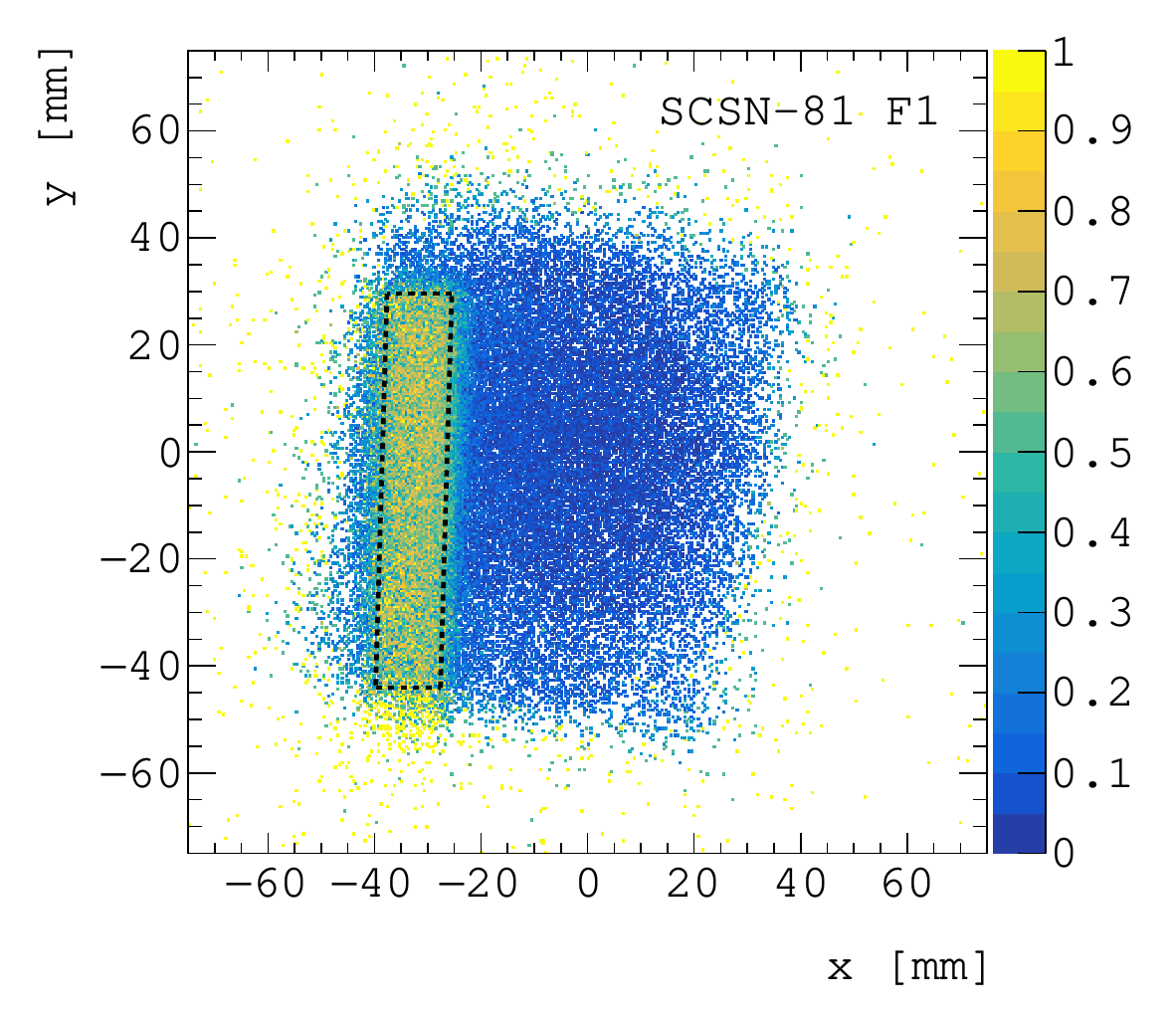}%
    \includegraphics[width=0.485\textwidth]{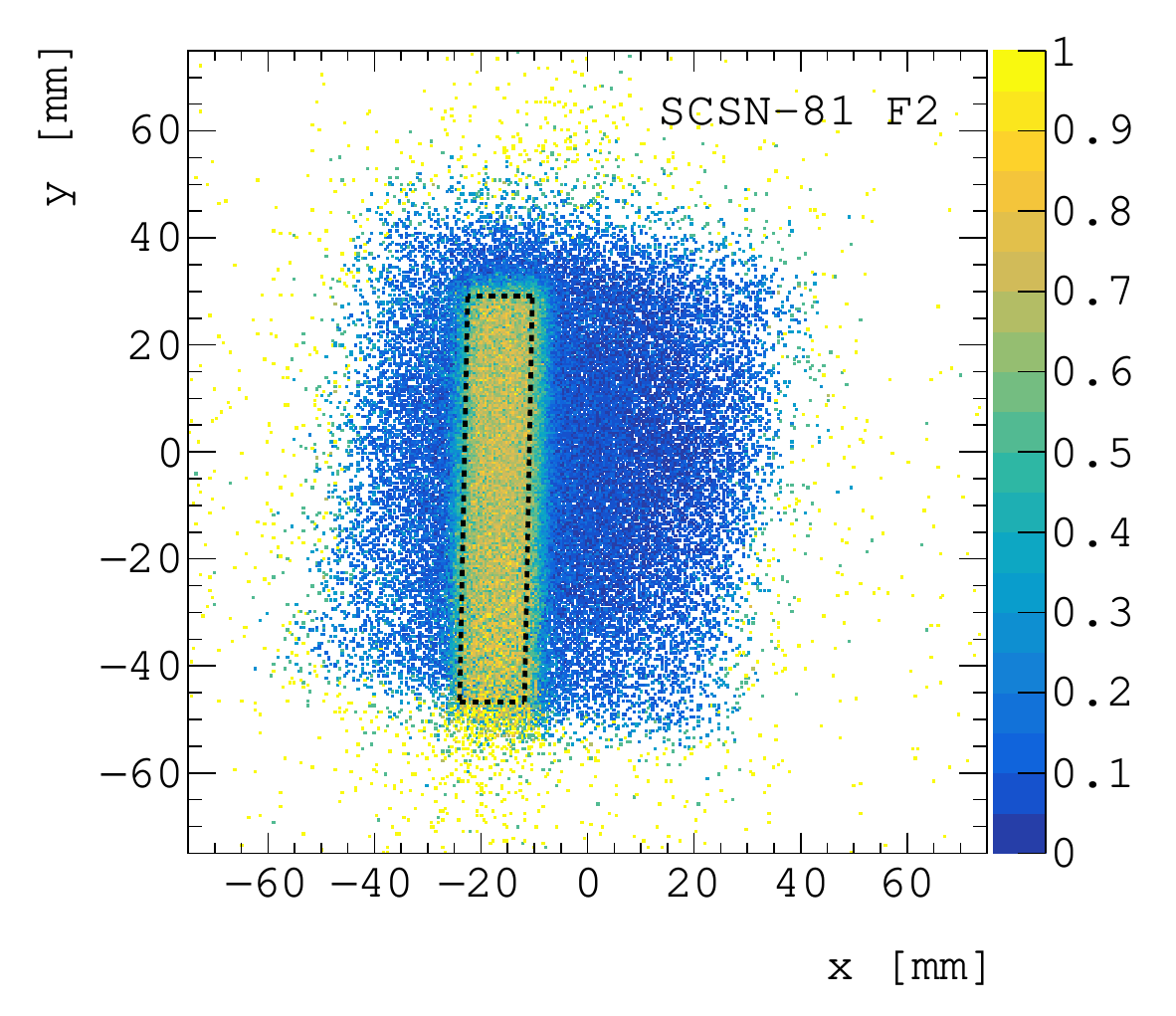}
    \includegraphics[width=0.485\textwidth]{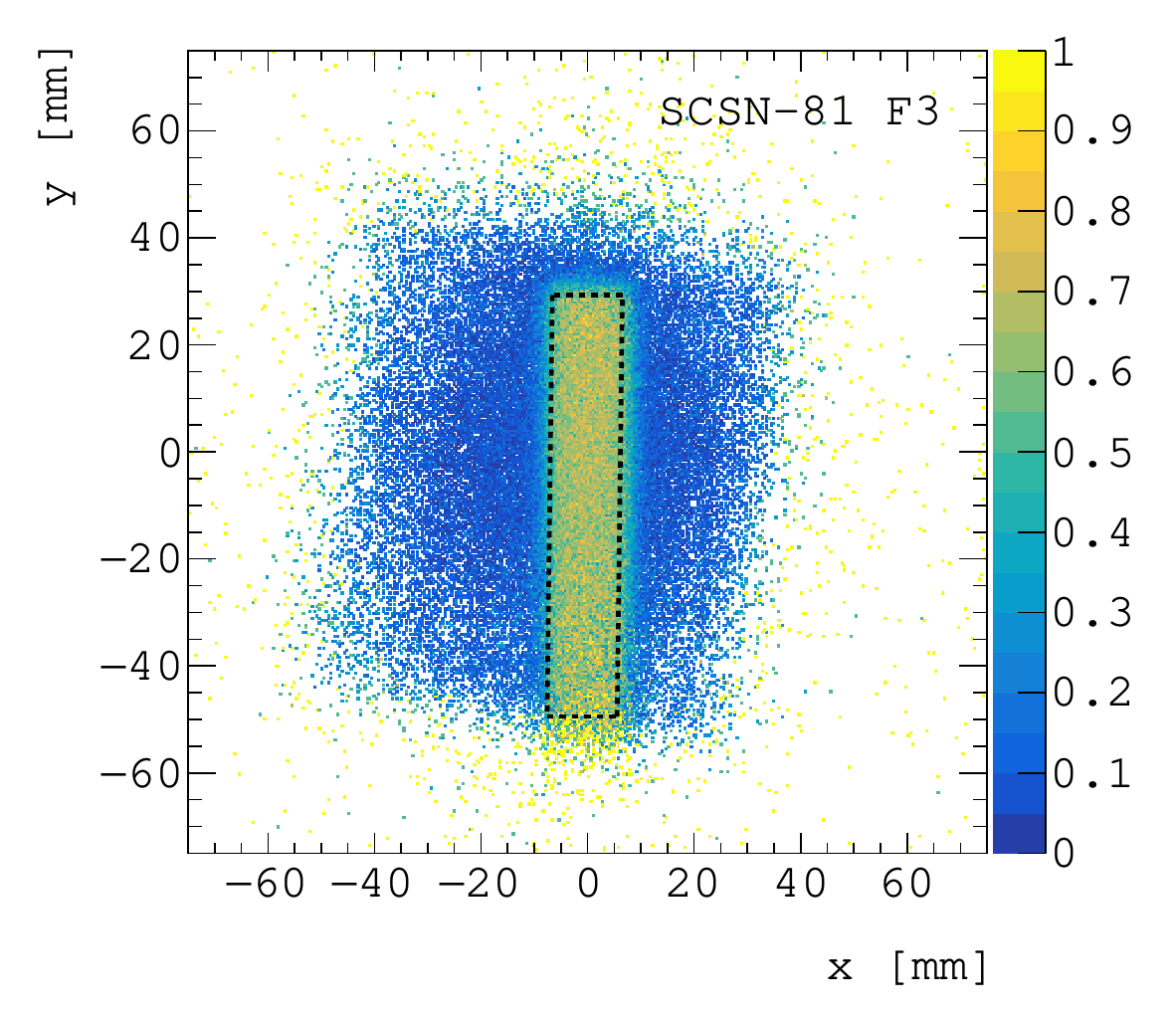}%
    \includegraphics[width=0.485\textwidth]{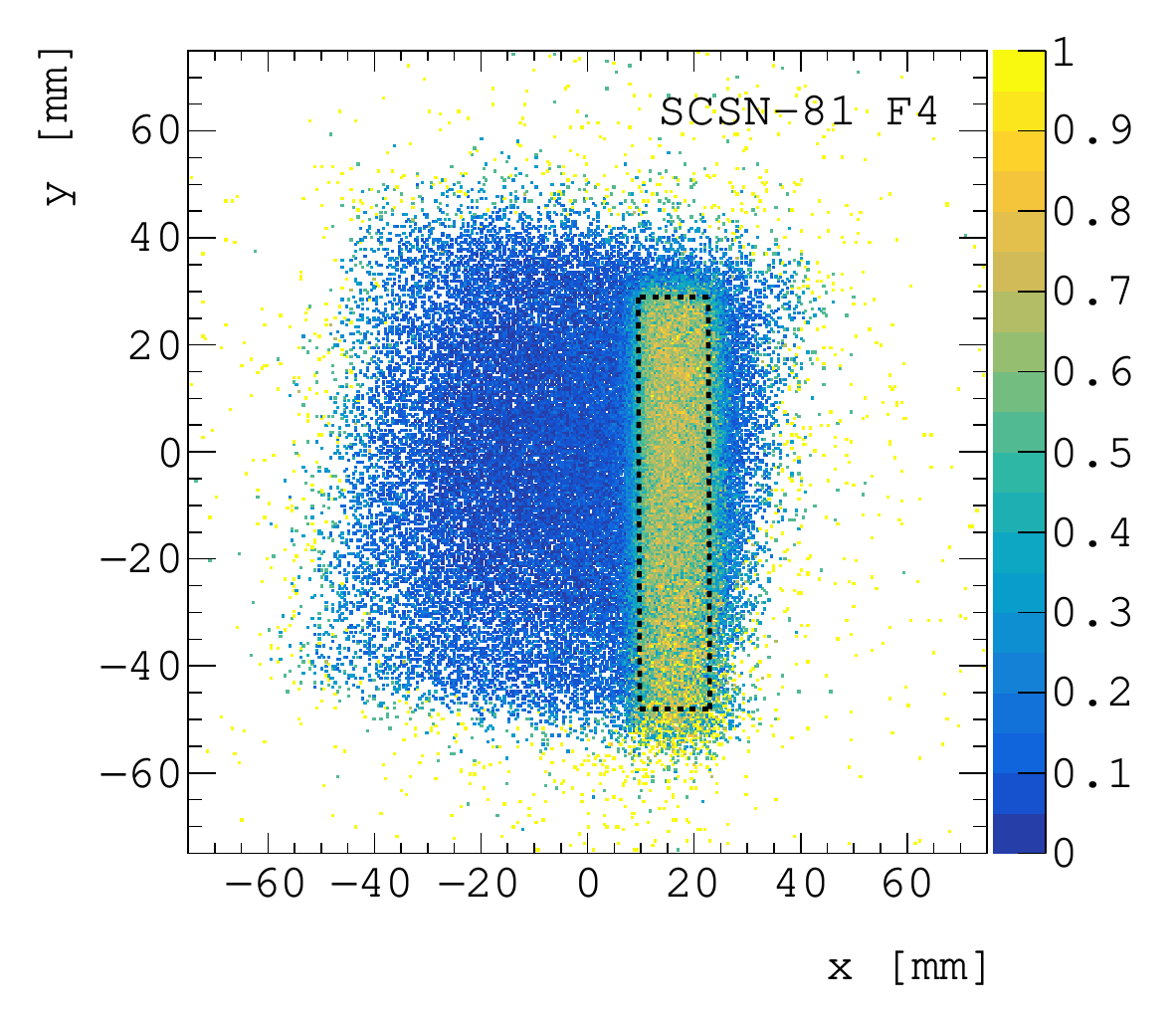}
    \caption{Efficiency maps for the finger tiles studied. The dashed
      lines indicate the fiducial region that corresponds
      approximately to the overlap between the tile area and the beam,
      positioned along a path in which the hit efficiency, determined
      from the ratio of hits with an integrated pulse $>25\unit{fC}$,
      divided by all the hits in the same bin, is $>50\%$.}
    \label{fig:maps_finger}
  \end{center}
\end{figure}

For each type of tile, the 2D efficiency plots are shown in
Figs.~\ref{fig:maps_sigma} and~\ref{fig:maps_finger}. The $x$ and $y$
axes have been rotated so that the finger tiles point vertically, and
the sides of the $\sigma$ tiles are parallel to the new axes. The
efficiency plots do not show any significant dependence on $x$ or
$y$. The efficiency is defined in each bin to be the ratio of the
number of hits with an integrated pulse above 25\unit{fC}, divided by
the total number of hits. The dashed lines enclose the fiducial
region, for which the efficiency is at least $ 50\%$.

\begin{figure}[!ht]
  \begin{center}
    \includegraphics[width=0.485\textwidth]{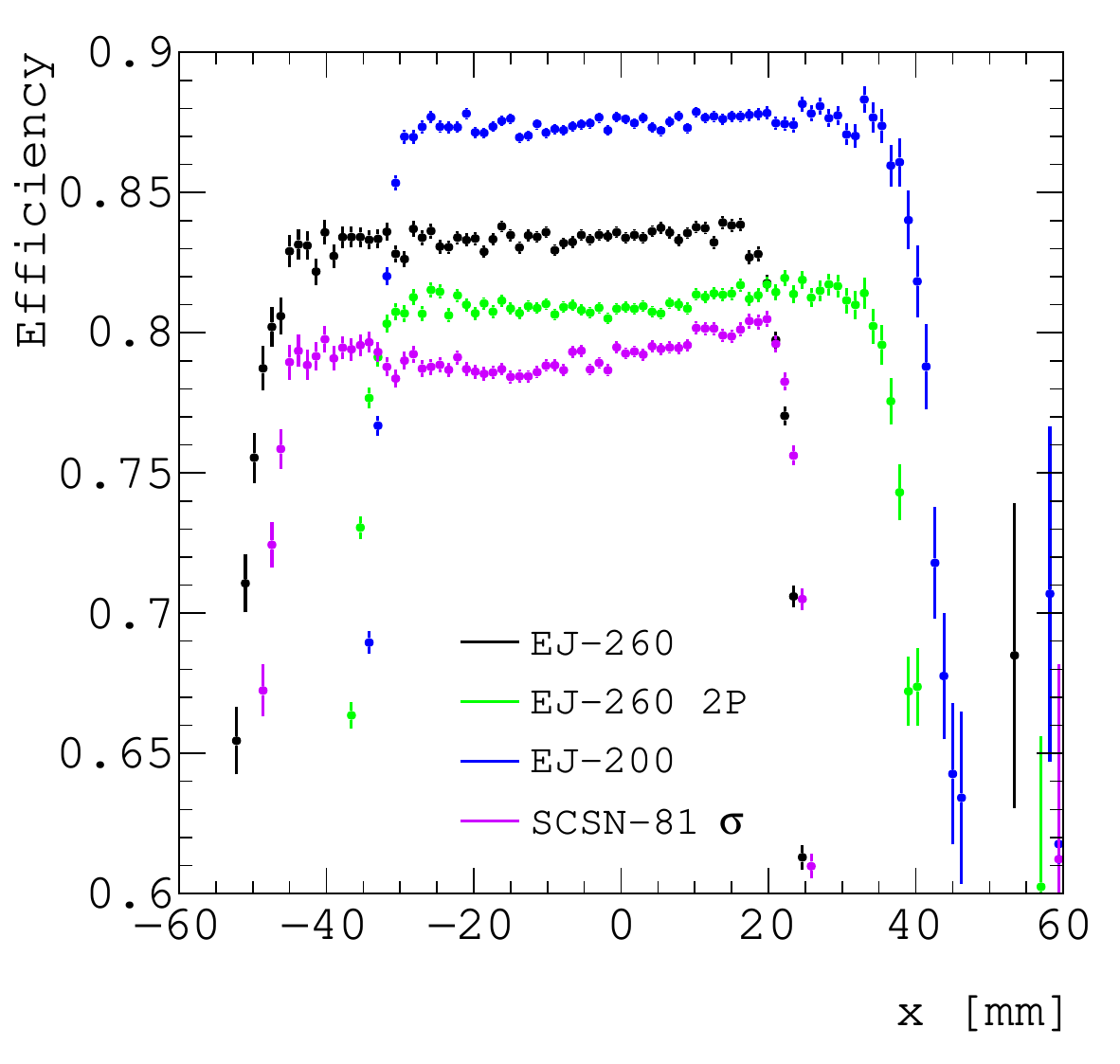}%
    \includegraphics[width=0.485\textwidth]{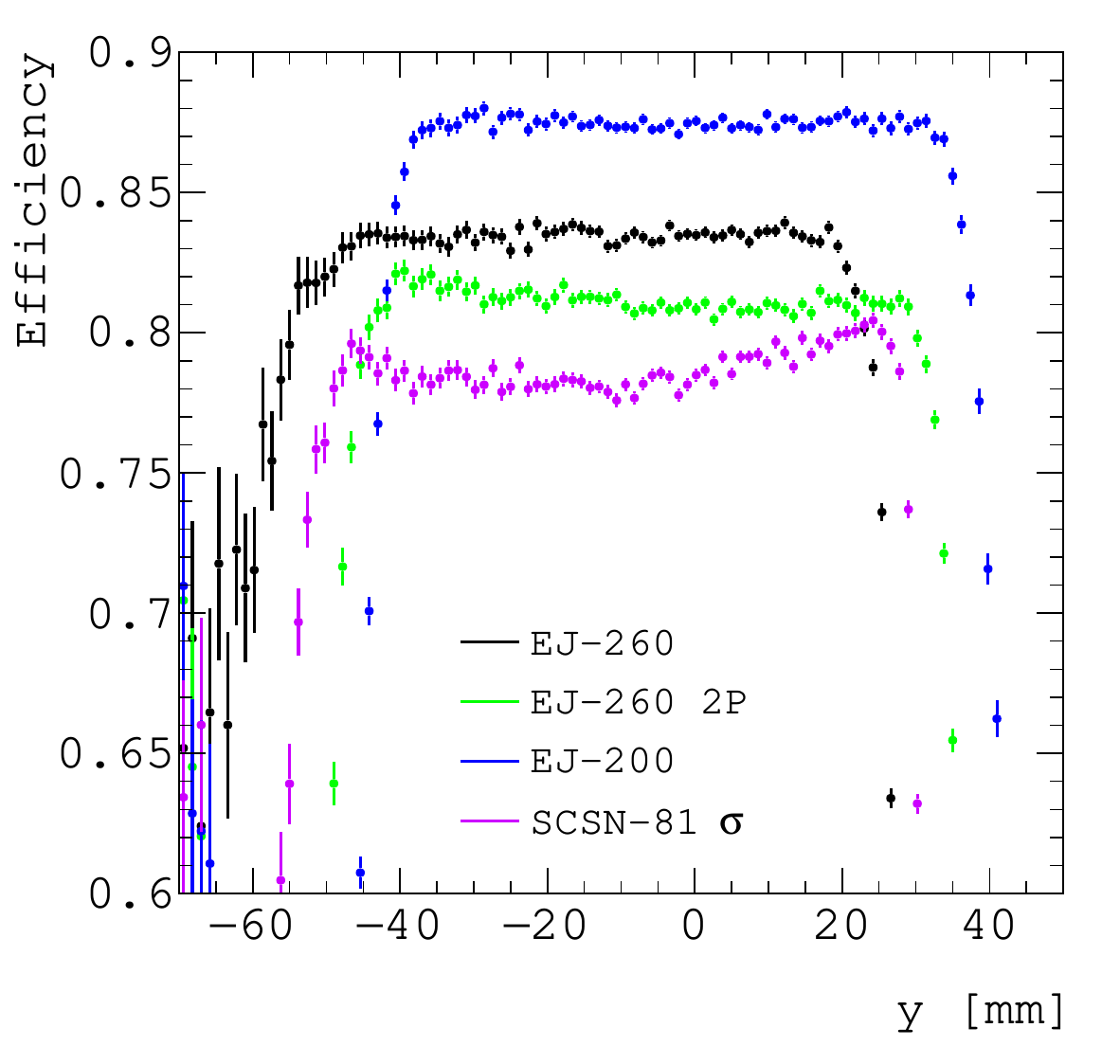}
    \includegraphics[width=0.485\textwidth]{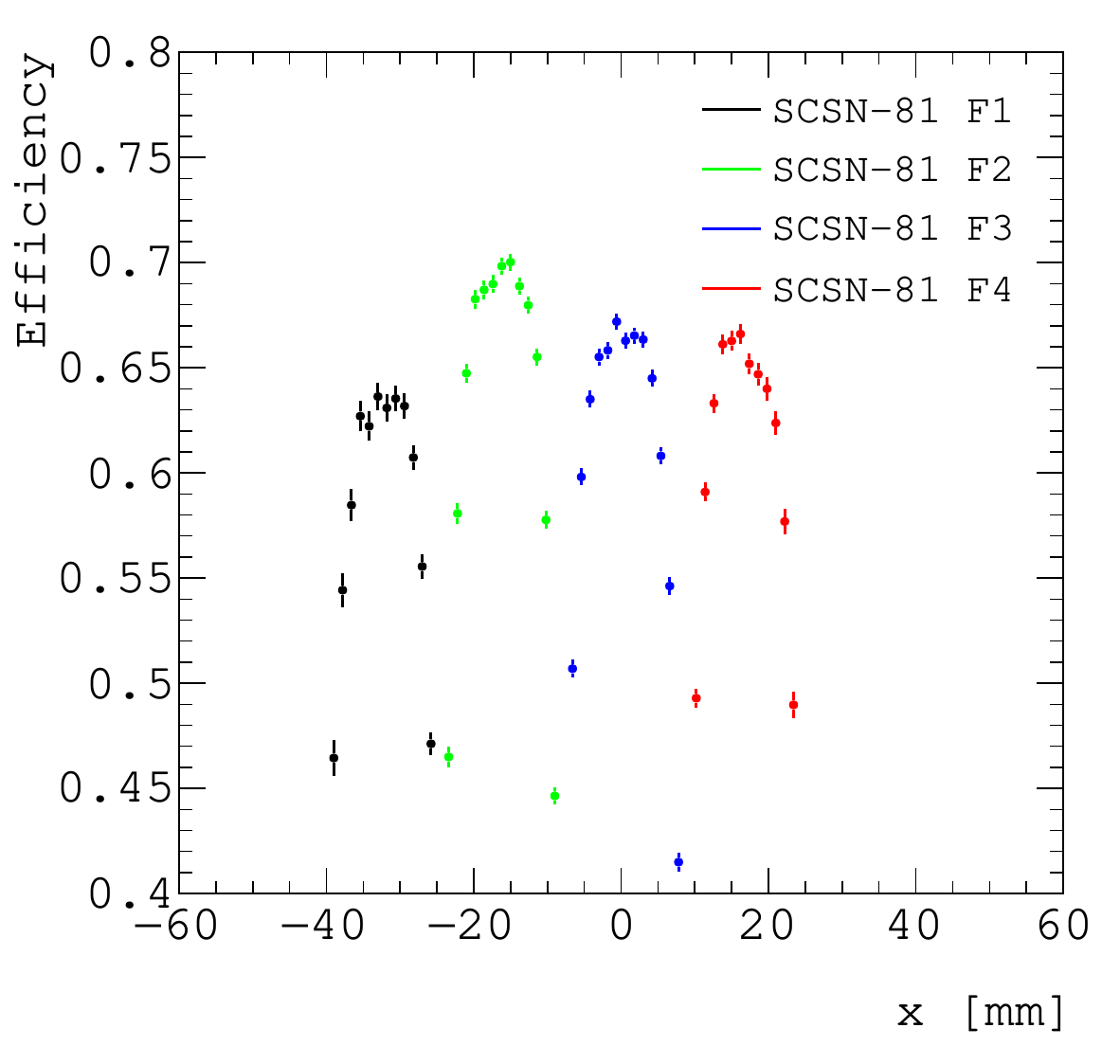}%
    \includegraphics[width=0.485\textwidth]{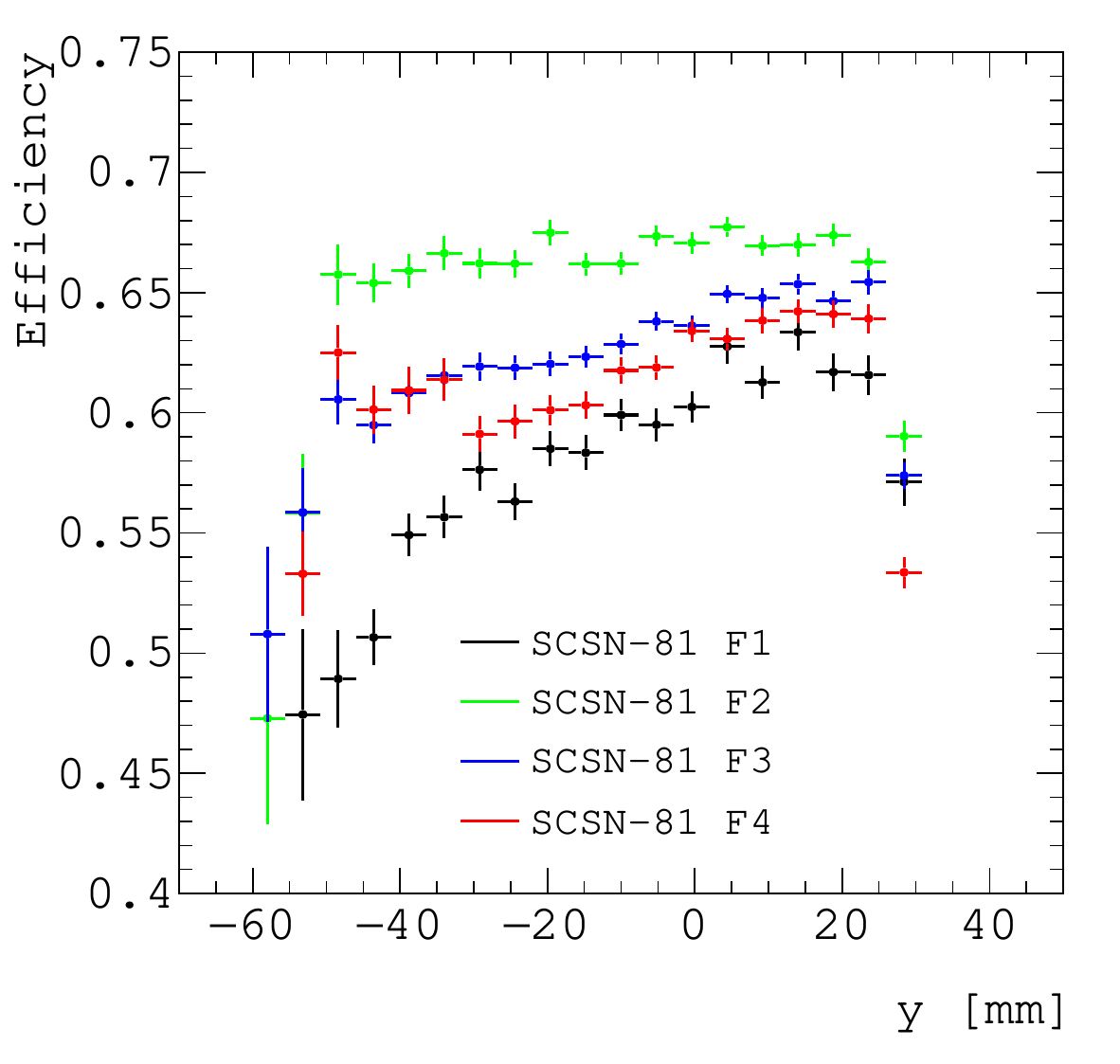}
    \caption{$x$ (left) and $y$ (right) efficiencies for $\sigma$ tiles
    (top) and finger tiles (bottom).}
    \label{fig:XandYeff}
  \end{center}
\end{figure}

To enhance any effects of non-uniformity in efficiency, possibly caused
by radiation damage in the irradiated tiles, projections of the 2D
efficiency maps along the $x$ and $y$ axes are shown in
Fig.~\ref{fig:XandYeff}. In fact, a small decrease in efficiency in
the middle section of the $\sigma$ tiles can be seen in the region
farthest from the wavelength-shifting fibers used to collect the
scintillation light. The finger tiles display a clear reduction in the
light yield the farther away the hits are from the top of the
scintillator tile. This suggests that the wavelength-shifting fiber
may be damaged, as the light produced in the lower section of the
finger tile has to travel a longer distance along that
fiber. Measurements at the University of Maryland using a collimated
Sr-90 source confirmed the trends observed in the test-beam data.

\subsection{Light yield}
\label{sec:light-yield_measurement}

The light yield in each tile is a parameter of interest, which, in
addition to the hit efficiency, characterizes the performance of a
scintillator. The light yield is defined as the average number of
photons collected in the tile from each incident muon. Each muon
releases approximately the same amount of energy in each tile. As
indicated in Table~\ref{tab:scintillators}, the Eljen tiles are
$4\unit{mm}$ thick, while the SCSN tiles are $3.7\unit{mm}$
thick. Hence, it is expected that a MIP deposits $\approx 9\%$ more
energy while traversing an Eljen tile than a SCSN one; this aspect
should be taken into consideration when comparing the light yields
among the different scintillator tiles.

The charge distributions present a set of peaks at a fixed distance
relative to each other, as is clearly seen, for example, in
Fig.~\ref{fig:multigauss_fit}.
%As mentioned previously, the charge distribution in, e.g.,
%Fig.~\ref{fig:energy_nofid} shows a set of peaks at a fixed distance
%relative to each other.
Each subsequent peak corresponds to a higher number of photons. We
discuss below two methods used to obtain a robust estimation of the
light yield.

%The physical processes that produced the entries for these histograms
%is well known. An incident muon causes the plastic material to
%scintillate, and the energy deposited follows a Landau
%distribution. From this energy, an integer number of photons is
%created, and so Landau distribution is used as the mean of a Poisson
%distribution. This produces a series of peaks with equal separation,
%the height of which follows a Landau distribution if connected. The
%photons are collected by the wavelength shifting fibers in the tiles
%and are sent to the read out module, where a Silicon Photomultiplier
%(SiPM) converts the optical data to a digital signal. For each photon
%that the SiPM receives, it produces a Gaussian distribution centered
%on the mean of the energy of the photon. Each peak produced by the
%Poisson distribution is therefore ``smeared'' by a Gaussian.

\subsubsection*{Method I -- Integration of contributions}

The charge distribution shows a series of equally spaced Gaussian
contributions of similar widths. Assuming that these contributions
have the same separation, we can estimate the average number of
photoelectrons by dividing the integrated signals of the charge
spectrum by the distance between neighboring peaks, starting at
$25\unit{fC}$ to eliminate the pedestal peak, and defining thereby the
mean signal produced by just one photoelectron as the corresponding
distance between neighboring peaks. We estimate this to be $\approx
41\unit{fC}$.

\begin{figure}[!h]
  \begin{center}
    \includegraphics[width=0.485\textwidth]{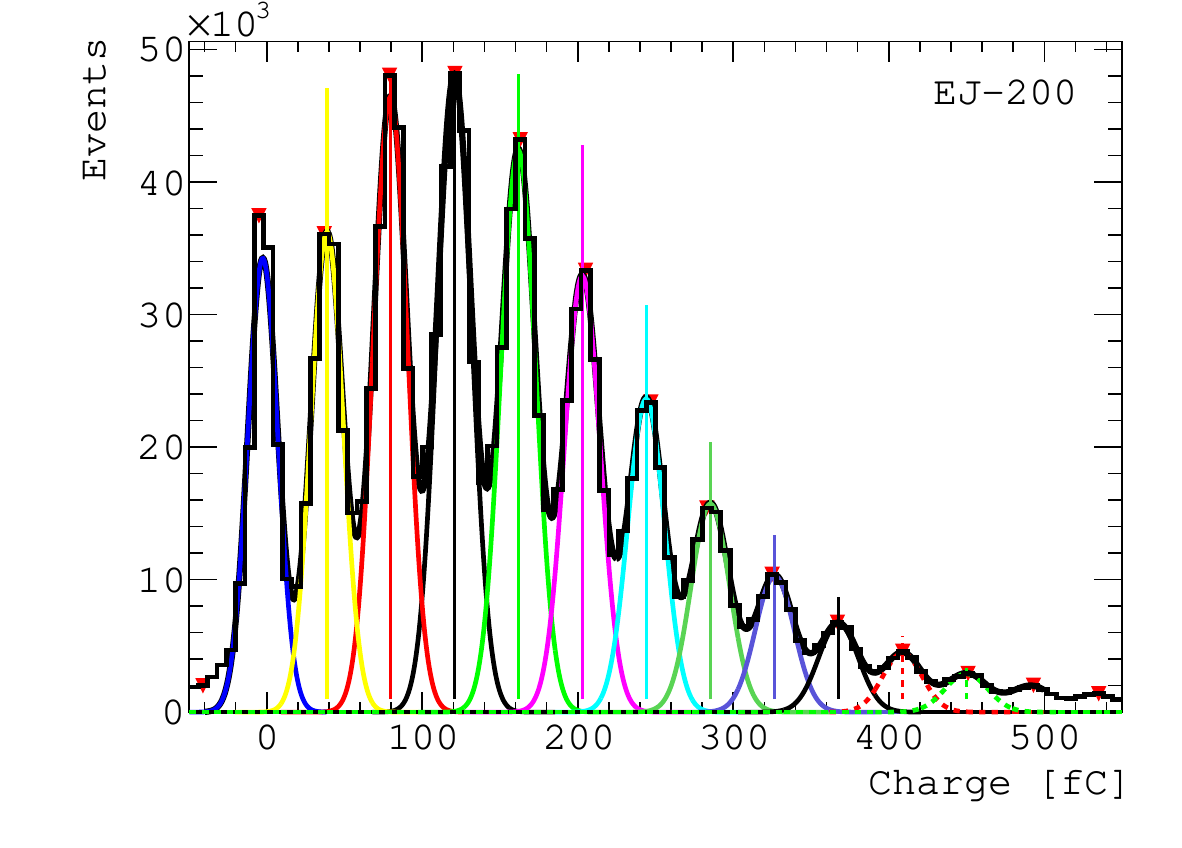}%
    \includegraphics[width=0.485\textwidth]{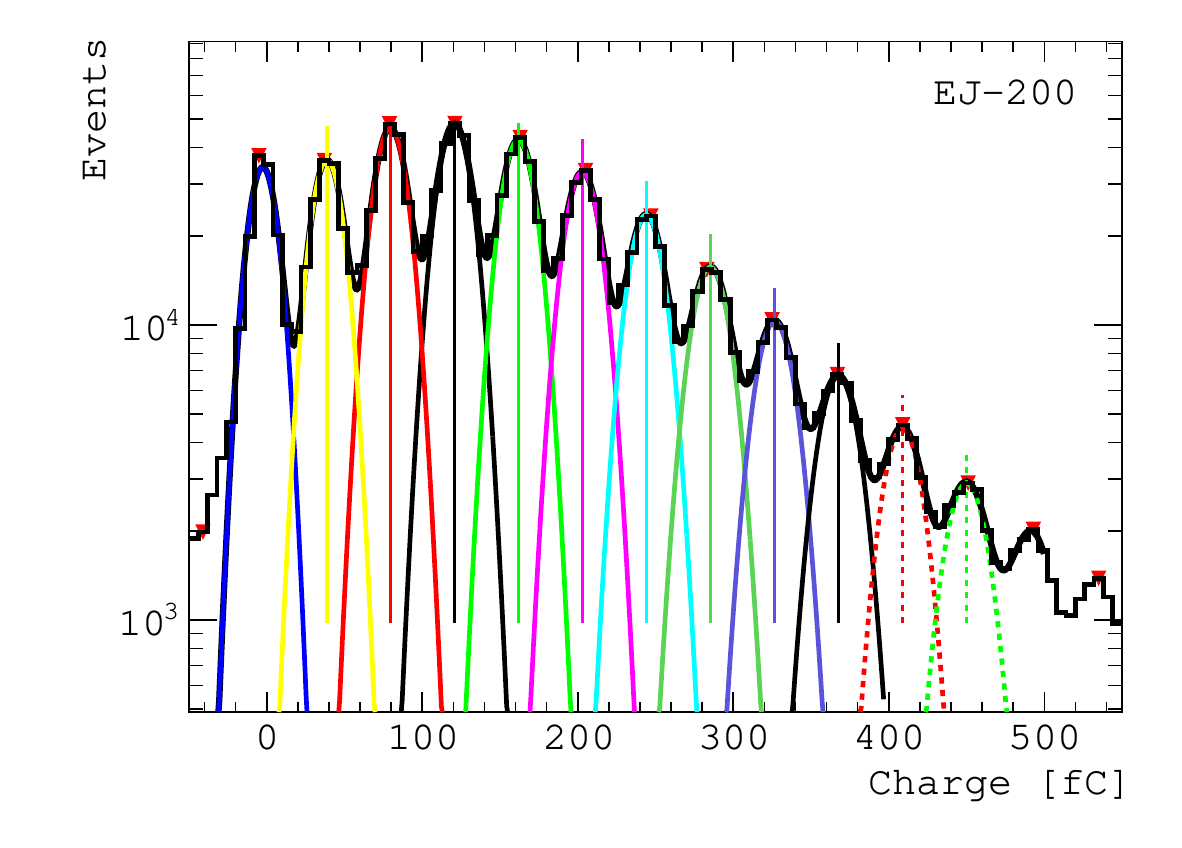}
    \includegraphics[width=0.485\textwidth]{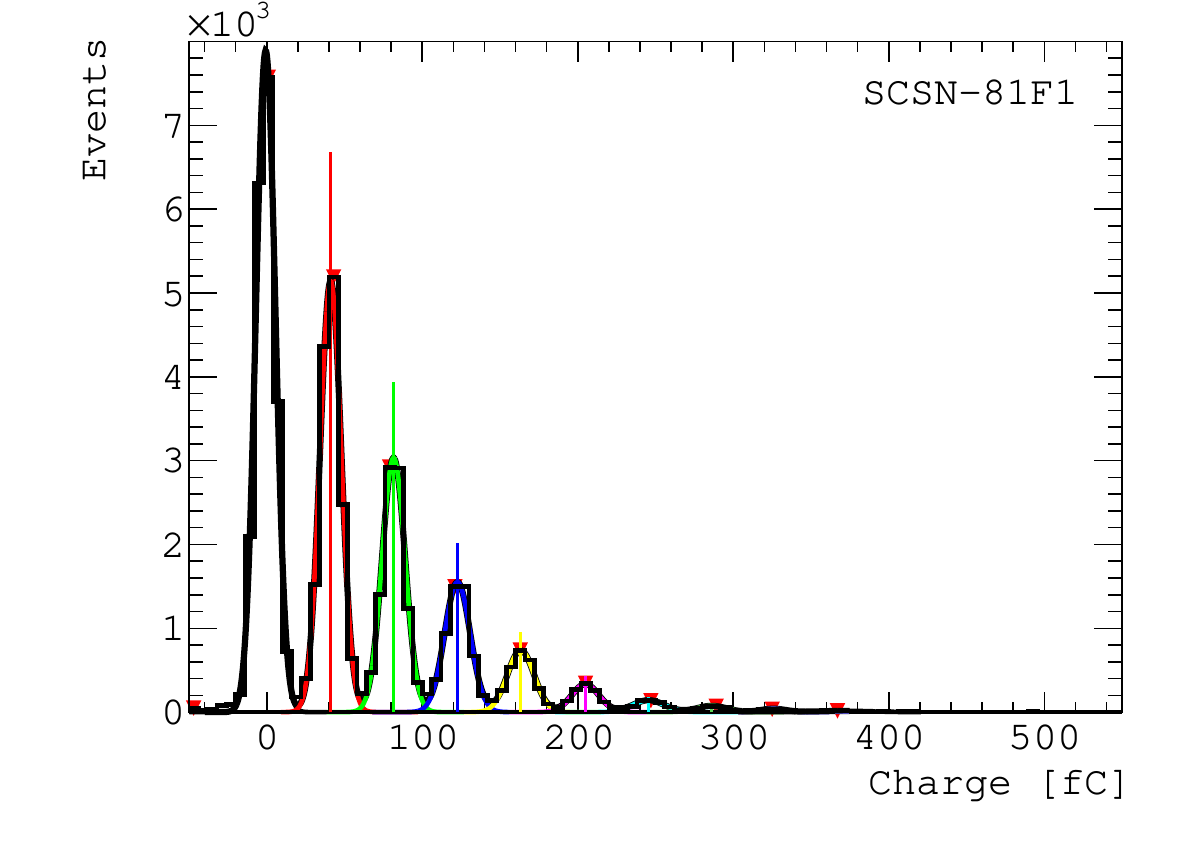}%
    \includegraphics[width=0.485\textwidth]{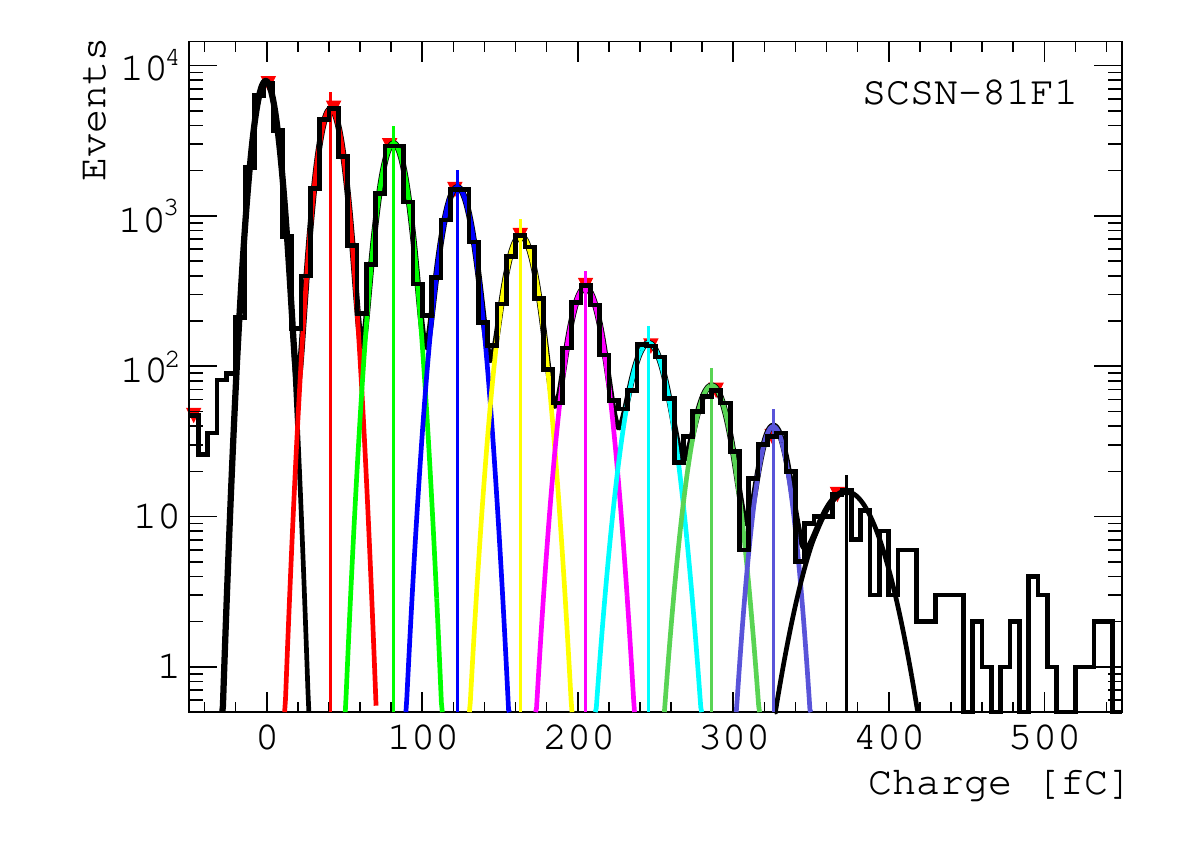}
    \caption{Example of multi-Gaussian fits to the charge spectra for
    the EJ-200 tile (upper), and the first SCSN-81 finger tile
    (lower), using linear (left) and logarithmic (right) scales. The
    colored Gaussians represent the fitted results; the vertical lines
    indicate the Gaussian functions included in the calculation of the
    estimator of the average number of photoelectrons.}
    \label{fig:multigauss_fit}
  \end{center}
\end{figure}

A simple fit of the charge distribution using a sum of Gaussian
functions, covering the whole spectrum, provides a cross check of the
result obtained.
%The initial values of the parameters are estimated using
%the ROOT~\cite{Brun:1997pa} \verb+TSpectrum+ class, which identifies
%the location and amplitude of the peaks in the charge spectrum. Each
%amplitude provides an initial value for the normalization of the
%corresponding Gaussian, by assuming a Gaussian width of $10\unit{fC}$.
%After the fit converges, the estimated average number of
%photoelectrons $\avgpe$ is obtained as follows:
%
%\begin{equation}
%\avgpe = \frac{\sum_{i=1}^{N}i\cdot A_i}{\sum_{i=1}^{N}A_i}\,,
%\end{equation}
%
%where $A_i$ is the normalization of the Gaussian corresponding to the
%$i^{th}$ peak in the energy spectrum.
Two examples of the fitted spectra are presented in
Fig.~\ref{fig:multigauss_fit}. The two representative fits capture the
main features of SiPM spectra: the distance between neighboring
Gaussian peaks is consistent with a constant value; the width of the
Gaussian peaks monotonically increases as they correspond to a higher
number of photoelectrons. The latter is not the case for the last
three Gaussians in the SCSN-81 finger-tile sample. Those Gaussians
count for less than 1\% of the number of events, indicating that the
simplified fit model is suitable to extract approximate estimates of
the scintillator performance.

It is to be noted, however, that the multi-Gaussian fit underestimates
the yields relative to the integral-of-contributions method by
$\approx 1-5\%$, partially because of the tail at high charge, where
it is not possible to fit additional Gaussian functions properly. The
comparison indicates that the integral-of-contribution method is
robust.

\subsubsection*{Method II -- Functional-form fit}

\begin{figure}[!h]
  \begin{center}
    \includegraphics[width=0.485\textwidth]{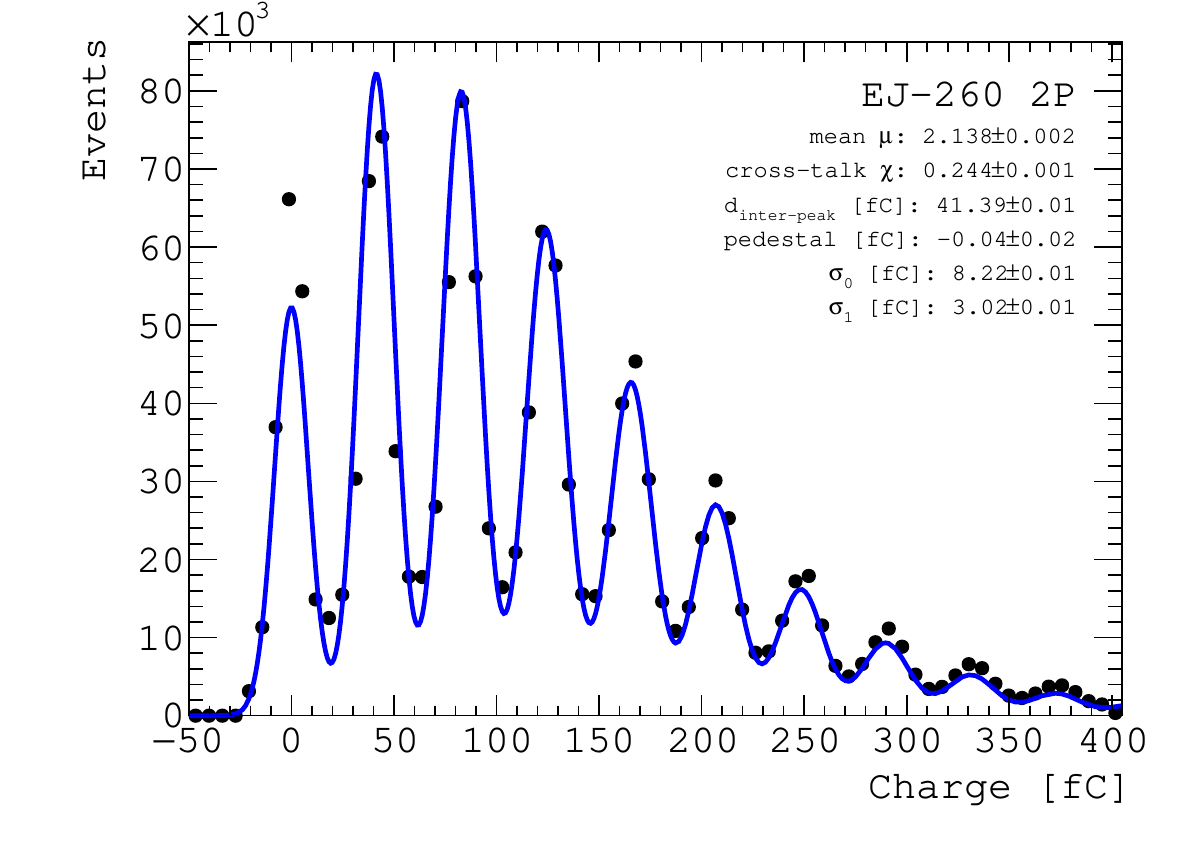}%
    \includegraphics[width=0.485\textwidth]{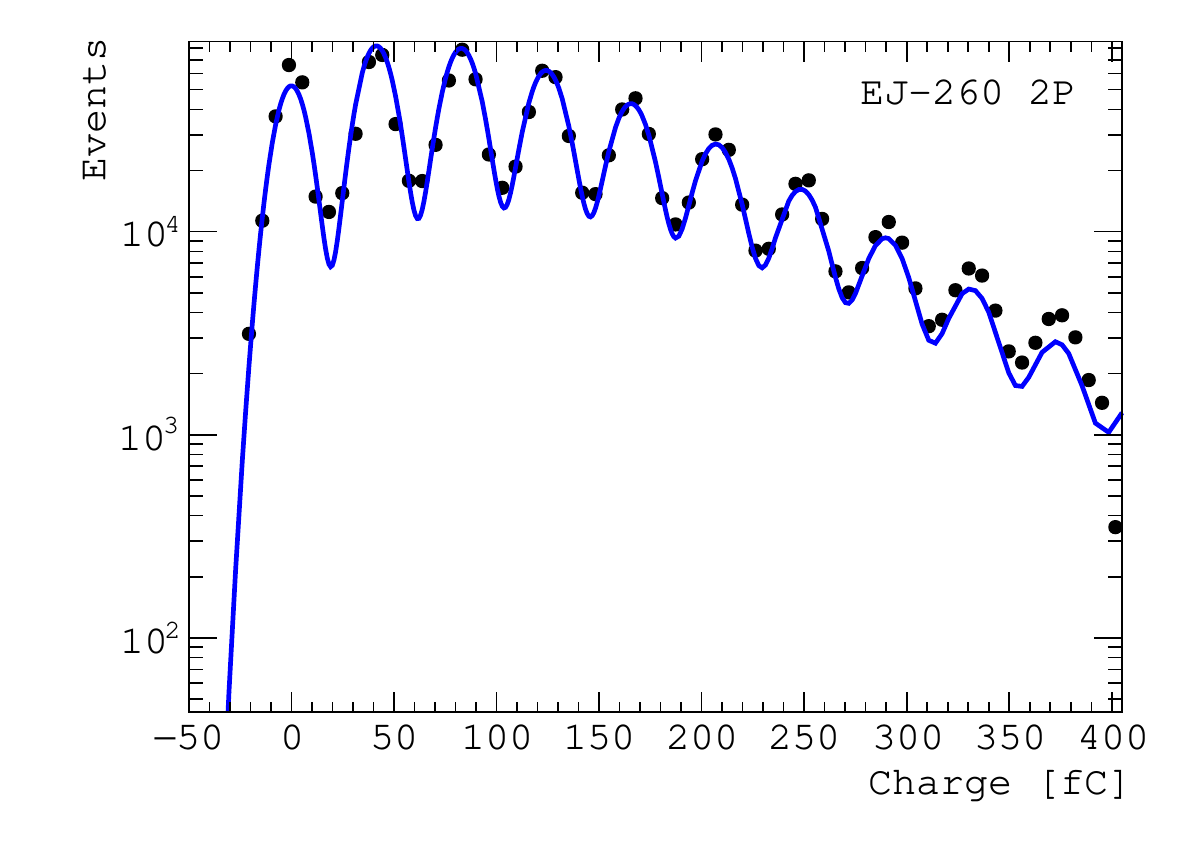}
    \includegraphics[width=0.485\textwidth]{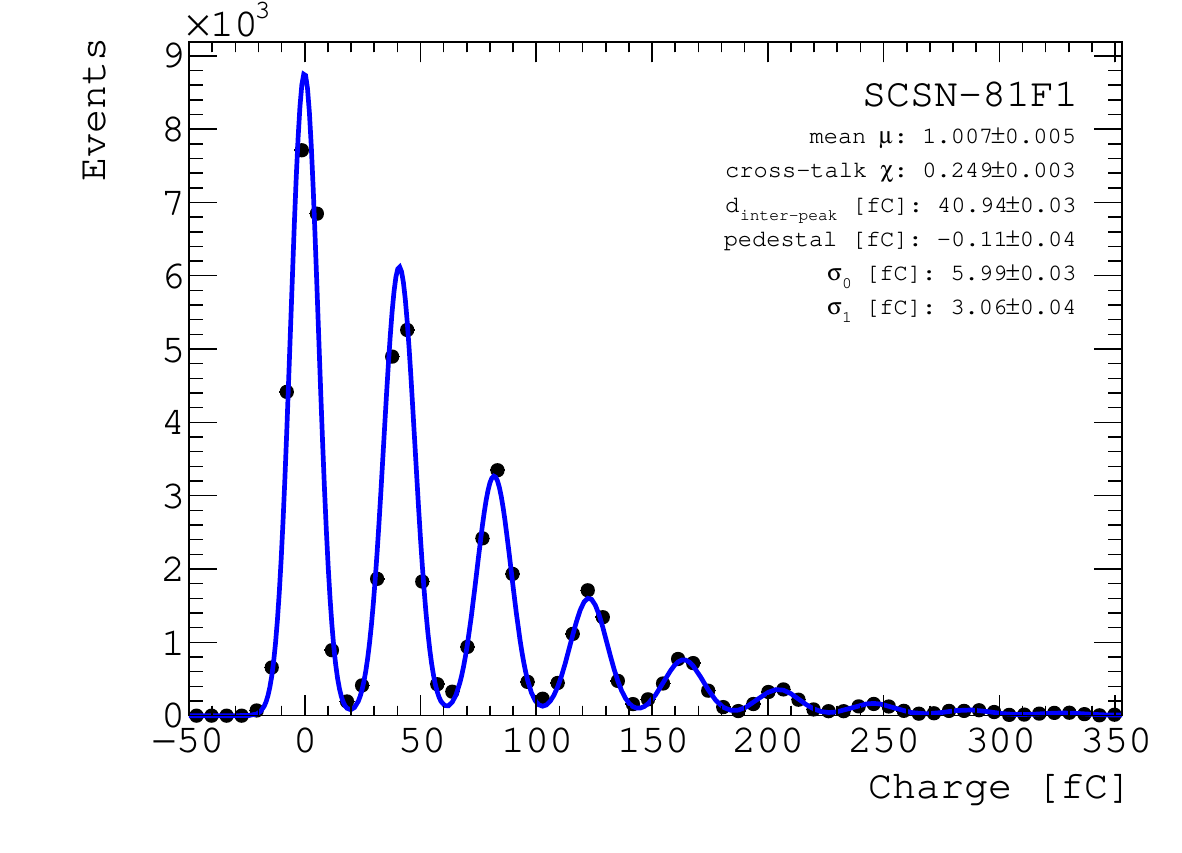}%
    \includegraphics[width=0.485\textwidth]{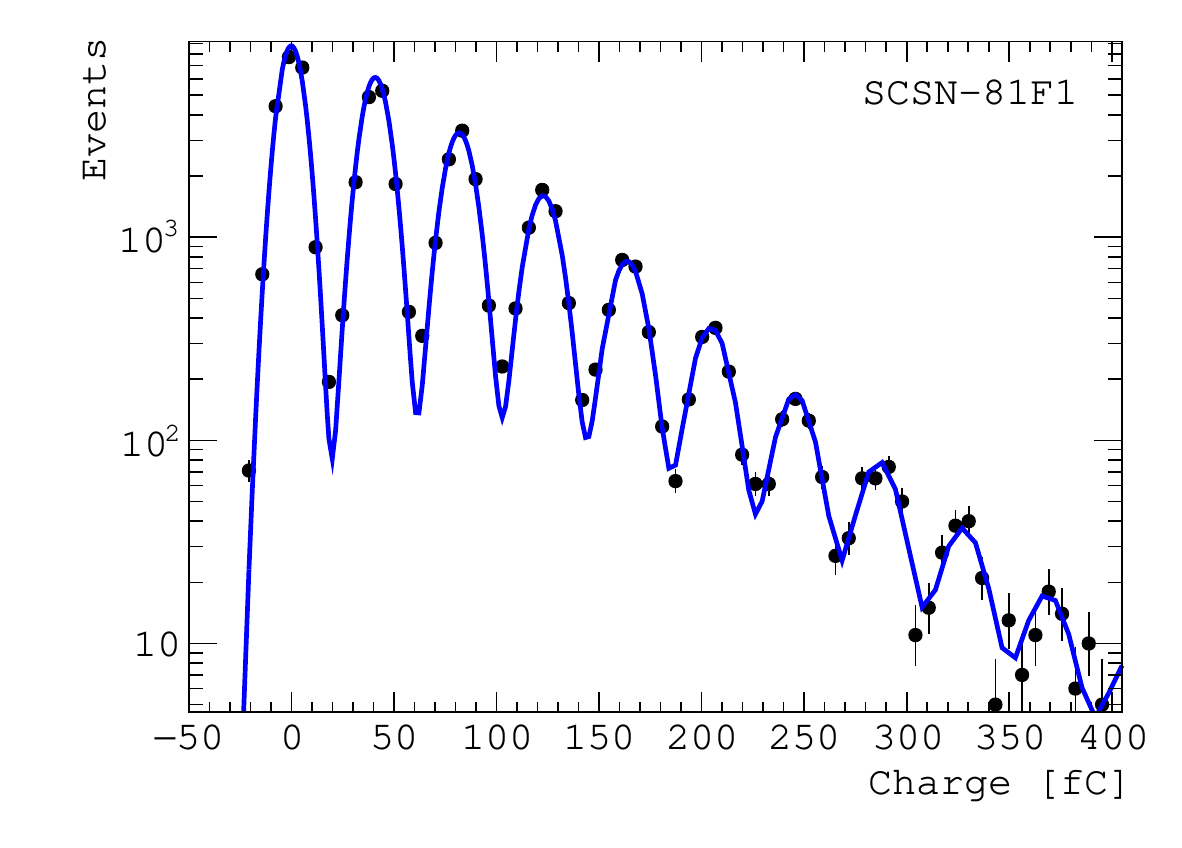}
    \caption{Example of functional-form fits to charge spectra for the
      EJ-260 2P tile (upper), and first SCSN-81 finger tile (lower),
      using linear (left) and logarithmic (right) scale. The fit
      parameters are reported in the linear-scale plot: $\mu$ is the
      average number of photons initiating the Geiger discharge in the
      SiPM; $\chi$ the probability of cross-talk;
      $d_{\mathrm{inter-peak}}$ is the separation between neighboring
      peaks; pedestal indicates the location of the pedestal peak (it
      is consistent with zero, suggesting that the pedestal
      subtraction was correctly implemented); $\sigma_0$ and
      $\sigma_1$ are used to parameterize the width of the Gaussian
      peaks as follows: $\sigma_i^2 = \sigma_0^2+i\sigma_1^2$.}
    \label{fig:physics_fit}
  \end{center}
\end{figure}

The second method is based on a fit to the charge spectra using a
function presented in Ref.~\cite{Chmill:2016ghf}. The RooFit
framework~\cite{Verkerke:2003ir} defines the probability density
function that is used to perform an unbinned maximum likelihood
fit. The fit parameters include the average number of initially
emitted photo-electrons, the distance between neighboring peaks, and
the cross-talk probability. It is important to note that the average
number of photoelectrons returned by the fit is not compared directly
with the number obtained with the other method described above; the
latter include all the secondary emissions. As a first-order
approximation, the results of the first method are compared to the
number of initially emitted photoelectrons divided by $1-\chi$, where
$\chi$ is the cross-talk probability. The fit also estimates the
distance between neighboring peaks. In all cases, this distance lies
within the interval $41.5\pm1.5\unit{fC}$, which is very close to the
value of $41\unit{fC}$ used to estimate the number of photoelectrons
via the integration of contributions. The SiPM bias voltage was indeed
set to obtain a distance between neighboring peaks of
$41\unit{fC}$. Thus, this result is in excellent agreement with the
configuration of the photo sensors. A fit including a probability
distribution that accounts for the effect of the after-pulsing has
also been performed; the result suggests that after-pulsing is
negligible. Two examples of the functional-form fits of the charge
spectra are shown in Fig.~\ref{fig:physics_fit}. The fits have only
six free parameters: pedestal peak location, separation between
neighboring peaks, electronic noise standard deviation, contribution
to the peak width dependent on the number of photoelectrons, average
number of photoelectrons, cross-talk probability. They demonstrate
that it is possible to obtain an adequate description of a SiPM
spectrum over multiple orders of magnitude with a small number of free
parameters.

\subsubsection*{Results}

Table~\ref{tab:light_yields} summarizes the results of the two methods
described in the previous paragraphs. The results of the
functional-form fit, even after implementing the correction for
cross-talk photoelectrons, are $\approx 20\%$ and $40\%$ lower for
$\sigma$ and finger tiles than the estimators provided by the other
methods. It is hard to obtain a high-quality fit due to the complexity
of the fit function.

\sloppy To further characterize the decrease in efficiency in the central
region of the SCSN-81 $\sigma$ tile, the fiducial region is divided
into two parts: central (defined by the requirement that
${-30<x_\mathrm{hit}<0\unit{mm}}$ and ${-30<y_\mathrm{hit}<0\unit{mm}}$),
and peripheral (complementary to the central region). The
corresponding light yields are then measured separately in the two
regions, using the two methods presented above. The efficiency profile
suggests that the light yield should be smaller in the central region,
which is farther from the wavelength-shifting readout fiber. This is
indeed the case; though the observed drop of $\approx 3\%$ is expected
to have a negligible impact on the energy resolution of the
calorimeter.

% SCSN-81 central/peripheral
% fiducial:   2.958-3.014
% central:    2.890-2.944
% peripheral: 2.977-3.054

\begin{table}[h]
  \begin{center}
  \topcaption{Estimated number of photoelectrons emitted on average when a
    single minimum-ionizing particle (muon) crosses the scintillator
    tile. The parameter $\chi$ is the cross-talk probability, one of
    the parameters of the functional-form fit function. The quoted
    values are statistical uncertainties from the fit.}
  \label{tab:light_yields}
    \begin{tabular}{ l | c | c | c | c}
      Tile      & $\avgpe_{\substack{functional-\\form}}$ & $\chi$
                & \begin{tabular}{@{}c@{}}
                $\avgpe_{\substack{functional-\\form}}$\\$/\left(1-\chi\right)$
                \end{tabular}
                & $\avgpe_{\substack{integration\;of\\contributions}}$ \\
      \hline
      EJ-200    &$2.742\pm0.002$&$0.235\pm0.001$&$3.585\pm0.004$&4.365\\
      EJ-260    &$2.316\pm0.002$&$0.238\pm0.001$&$3.040\pm0.004$&3.784\\
      EJ-260 2P &$2.138\pm0.002$&$0.244\pm0.001$&$2.826\pm0.003$&3.531\\
      SCSN-81F1 &$1.007\pm0.005$&$0.249\pm0.003$&$1.340\pm0.009$&2.157\\
      SCSN-81F2 &$1.191\pm0.004$&$0.257\pm0.002$&$1.602\pm0.007$&2.410\\
      SCSN-81F3 &$1.085\pm0.003$&$0.252\pm0.002$&$1.451\pm0.006$&2.295\\
      SCSN-81F4 &$1.042\pm0.004$&$0.250\pm0.002$&$1.390\pm0.007$&2.240\\
      SCSN-81S  &$1.711\pm0.002$&$0.268\pm0.001$&$2.336\pm0.003$&3.026\\
    \end{tabular}
  \end{center}
\end{table}

\section{Summary and conclusions}
\label{sec:conclusion}

This study involves the measurement of the light yield and uniformity
of irradiated plastic scintillators using a muon beam produced from
the decay of a 150\GeV pion beam. The efficiency is measured relative
to the $x$ and $y$ positions within the scintillator tile detectors to
determine whether any color centers created during irradiation cause a
change in the tile efficiency. The irradiated SCSN-81 $\sigma$ tile is
found to have a considerable reduction, about $50\%$, in the light
yield relative to a similarly configured, unirradiated EJ-200
tile. This is expected, and consistent with studies presented in
Refs.~\cite{CMSHCAL:2016dvd,CMS:2020mce}. It is noted that the
radiation damage affects the uniformity of the light yield by at most
a few percent, with negligible impact on the resolution of the
calorimeter. A more significant dependence of the light yield on hit
position is observed in the case of finger tiles, which at worst
indicates an efficiency variation of about $40\%$.

%% file: content/acknowledgements.tex
The authors would like to thank CERN for the operations of the SPS
accelerator; Dragoslav Lazic for supporting operations in the H2
test-beam area; Chuck Hurlbut (Eljen Technology) for the advice on
scintillators, and the preparation of test samples; Janina Gielata
(FNAL) for the preparation of optical connections. Individuals have
received support from the Belgian Fonds de la Recherche Scientifique,
and Fonds voor Wetenschappelijk Onderzoek; the Brazilian Funding
Agencies (CNPq, CAPES, FAPERJ, FAPERGS, and FAPESP); SRNSF (Georgia);
the Bundesministerium f\"ur Bildung und Forschung, the Deutsche
Forschungsgemeinschaft (DFG), under Germany's Excellence Strategy --
EXC 2121 ``Quantum Universe'' -- 390833306, and under project number
400140256 - GRK2497, and Helmholtz-Gemeinschaft Deutscher
Forschungszentren, Germany; the National Research, Development and
Innovation Office (NKFIH) (Hungary) under project numbers K~128713,
K~143460, and TKP2021-NKTA-64; the Department of Atomic Energy and the
Department of Science and Technology, India; the Ministry of Science,
ICT and Future Planning, and National Research Foundation (NRF),
Republic of Korea; the Lithuanian Academy of Sciences; the Scientific
and Technical Research Council of Turkey, and Turkish Energy, Nuclear
and Mineral Research Agency; the National Academy of Sciences of
Ukraine; the US Department of Energy.

%% file: DN-18-007_authorlist_optC.tex
\parindent 0pt
\parskip 4pt
\newcommand{\cmsinstitute}[1]{\par\pagebreak[3]\bfseries #1 \mdseries\\[0pt]}
\newcommand{\cmsorcid}[1]{\href{https://orcid.org/#1}{\hspace*{0.1em}\raisebox{-0.45ex}{\includegraphics[width=1em]{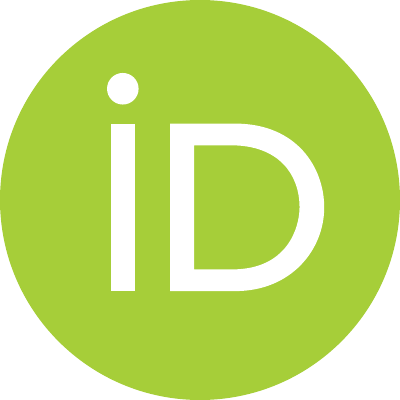}}}}
\makeatletter
\newcommand{\cmsAuthorMark}[1]{\hbox{\@textsuperscript{\normalfont#1}}}
\makeatother
\newskip{\cmsinstskip} \cmsinstskip=4pt

%\cmsinstitute{The CMS HCAL Collaboration}

\cmsinstitute{Yerevan~Physics~Institute, Yerevan, Armenia}
{\tolerance=6000
A.~Gevorgyan, A.~Petrosyan, A.~Tumasyan
\par}

\cmsinstitute{Universiteit~Antwerpen, Antwerpen, Belgium\cmsAuthorMark{*}}
{\tolerance=6000
M.~Van~De~Klundert, H.~Van~Haevermaet, P.~Van~Mechelen\cmsorcid{0000-0002-8731-9051}, A.~Van~Spilbeeck
\par}

\cmsinstitute{Centro~Brasileiro~de~Pesquisas~Fisicas, Rio de Janeiro, Brazil\cmsAuthorMark{*}}
{\tolerance=6000
G.A.~Alves\cmsorcid{0000-0002-8369-1446}, C.~Hensel\cmsorcid{0000-0001-8874-7624}
\par}

\cmsinstitute{Universidade~do~Estado~do~Rio~de~Janeiro, Rio de Janeiro, Brazil\cmsAuthorMark{*}}
{\tolerance=6000
W.L.~Ald\'{a}~J\'{u}nior\cmsorcid{0000-0001-5855-9817}, W.~Carvalho\cmsorcid{0000-0003-0738-6615}, J.~Chinellato\cmsAuthorMark{1}, C.~De~Olivera~Martins, D.~Matos~Figueiredo, C.~Mora~Herrera\cmsorcid{0000-0003-3915-3170}, H.~Nogima\cmsorcid{0000-0001-7705-1066}, W.L.~Prado~Da~Silva, E.J.~Tonelli~Manganote, A.~Vilela~Pereira\cmsorcid{0000-0003-3177-4626}
% ...
\par}

\vskip\cmsinstskip

\dag:~Deceased\\
$^{*}$No longer in CMS HCAL Collaboration\\
$^{1}$Also at Universidade Estadual de Campinas, Campinas, Brazil\\
% ...

%\end{document}